\documentclass[11pt,cite,epsfig,psfrag]{article}
\usepackage[utf8]{inputenc}
\usepackage{placeins}
\usepackage{geometry}
 \usepackage{graphicx}
 \usepackage{subfig}
  \usepackage{amsmath} 
  \graphicspath{{Images_isco/}}
  \usepackage[usenames, dvipsnames]{color}
\usepackage{wrapfig}
\usepackage{boxedminipage}
\usepackage{multirow}
\usepackage{xcolor}

\usepackage{fullpage}
\usepackage{amsmath}
\usepackage{amssymb}
\usepackage{graphicx}
\usepackage{mathrsfs}
\usepackage{wrapfig}
\usepackage{boxedminipage}
\usepackage{epsfig}
\usepackage{setspace}
\usepackage{float}
\usepackage{hyperref}
\usepackage{enumerate}
\usepackage{cite}
\usepackage[font=footnotesize,labelfont=rm]{caption}
\usepackage{soul}

\usepackage[numbers,sort&compress]{natbib}

\usepackage[normalem]{ulem}

\newcommand{\mul}{m}

\newcommand{\Or}{\mathcal{O}}

\newcommand{\non}{\nonumber}

\newcommand{\beq}{\begin{equation}}
	\newcommand{\eeq}{\end{equation}}
\newcommand{\beqa}{\begin{eqnarray}}
	\newcommand{\eeqa}{\end{eqnarray}}
\newcommand{\beqar}{\begin{eqnarray*}}
	\newcommand{\bal}{\begin{align}}
		\newcommand{\eal}{\end{align}}	

\def\half{{1\over 2}}
	\newcommand{\expect}[1]{ \langle #1 \rangle}

\def\be{\begin{equation}}
\def\ee{\end{equation}}
\def\bea{\begin{eqnarray}}
\def\eea{\end{eqnarray}}

\numberwithin{equation}{section}

 \newcommand{\RN}[1]{%
   \textup{\uppercase\expandafter{\romannumeral#1}}%
 }
\begin{document}

\thispagestyle{empty}

\vskip 2cm

\begin{center}
{\Large \bf Criticality of ISCOs and AdS/CFT}
\end{center}

\vskip .2cm


\begin{center}
{\bf Chandrasekhar Bhamidipati \footnote{chandrasekhar@iitbbs.ac.in},
\bf   Parashar Chatterjee \footnote{parasharchatterjee1@gmail.com},  \\
 Sudipta Mukherji \footnote{mukherji@iopb.res.in} and Yogesh Kumar Srivastava \footnote{yogeshs@niser.ac.in} }
\end{center}

\begin{center}{$^{1}$ Department of Physics, School of Basic Sciences\\ 
		Indian Institute of Technology Bhubaneswar \\ Bhubaneswar, Odisha, 752050, India}
\end{center}

\begin{center}{ $^{{2},{4}}$  National Institute of Science Education and Research (NISER), \\ Bhubaneswar, P.O. Jatni, Khurda, Odisha, 752050, India}	
\end{center}

\begin{center}{ $^{3}$ Physics and Applied Mathematics Unit, Indian Statistical Institute, Kolkata, India}	
\end{center}

\begin{center}{$^{{2},{4}}$ Homi Bhabha National Institute, Training School Complex, \\ Anushakti Nagar, Mumbai,  400085, India}	
\end{center}

\vskip 1.2cm
\centerline{\bf Abstract}
\vskip 0.5cm

We study the trajectories of massive particles in spherically symmetric black holes in arbitrary dimensions, and find certain universal features based on the topological classification of the fixed points. If the system admits a center, we find two possible outcomes: regardless of
the value of the angular momentum, the center always survives, which is realized in global AdS spacetimes or, the center disappears below a critical value of angular momentum, which happens for various spherically symmetric black holes. For the latter case, we find that irrespective of the details of the black hole, there must always be a saddle point. Topological arguments show that there exists a certain critical value of energy, angular momentum and the angular velocity, where the center and the saddle coalesce. This happens at a special point in the parameter space where the trajectories are the limiting innermost stable circular orbits (ISCOs). At the critical point, conserved quantities show universal, van der Waals-like mean-field scaling typical of a second-order phase transition. The anomalous dimensions $\gamma$ of the double-twist operators in the CFT are found, both using AdS/CFT and through the the heavy-heavy-light-light four point correlators, giving negative and positive values for the center and saddle, respectively, including the emergence of certain non-analytic behaviour at the ISCO. For the center, we also find subleading corrections in $\frac{1}{\Delta_H}$ to $\gamma$ in the dual CFT, and discuss the implications of our results. 

\noindent

\vskip 0.5cm
\noindent

\newpage
\setcounter{footnote}{0}
\noindent

\baselineskip 15pt
\section{Introduction}

\medskip
\noindent
In theories of gravity, the geodesic of a test particle encodes fundamental information about the  background geometry. For instance, the innermost stable circular orbit (ISCO) of a timelike geodesic 
around a stationary black hole serves as a proxy for its size, while  for a rotating
 black hole, the ISCO offers a window into the spin of the black hole (for a review, see~\cite{Narayan:2005ie}). Therefore, considerable research has focused on deriving exact solutions of the geodesic equations or analyzing qualitative behavior emanating from the equations  in the absence of exact solutions. \\
 
 \noindent
 For a stationary spherically symmetric black hole in four or higher dimensions,  timelike geodesics lying on the equatorial plane are parametrized by the  conserved energy and the angular momentum of the particle. 
 As one or both of these parameters are varied, one finds either periodic orbits or trajectories that plunge into the black hole. Since  the geodesic equation can  generally be cast into a pair of first order non-linear differential equations, a systematic approach 
 to examining them is to carry out a phase-plane analysis.  Consider for example a massive particle in four dimensional Schwarzschild spacetime. The relevant equation for
 the geodesic can be written in the form
 \begin{equation}
\Big( \frac{dx}{d\phi}\Big)^2 = \frac{E^2}{L^2}  -\frac{1}{L^2}(1-2 \mul x) (1 + L^2 x^2).\nonumber
\end{equation}
Here  $x$ is the inverse of the radial  distance measured from the singularity of the hole and $\phi$ is the azimuthal coordinate. While the quantities $E$ and $L$ of the above equation  are the energy and the angular momentum of the particle, the parameter $m$ is related to the mass of the black hole. Taking a derivative with respect to $\phi$ and defining $dx/d\phi = z$, the second order equation can be recast in the following form ~\cite{Dean:1999jr}\begin{eqnarray}
&&\frac{dx}{d\phi} = z,\nonumber\\
&&\frac{dz}{d\phi} = \frac{\mul}{L^2}(1 + x^2 L^2)  - x (1 - 2 \mul x).\nonumber
\end{eqnarray}
Qualitative behaviors of the geodesics are now straightforwardly obtained. Firstly, the fixed points follow from the equations $dx/d\phi = dz/d\phi = 0$. Upon solving, we get
\begin{equation}
z = 0, ~~x_{fp\pm} = \frac{1}{6\mul L}\Big(L \pm \sqrt{L^2 - 12 \mul^2}\Big).\nonumber
\end{equation}
Here the subscript $fp$ on $x$ indicates that they are the solutions of the fixed point equations. Secondly, a linear stability analysis around these fixed points tells us  that
$x_{fp-}$ and $x_{fp+}$ represent a center and a saddle  respectively.  It follows immediately from the solutions  that, as we reduce the angular momentum of the particle,  two fixed points approach one another and for  $L^2 = 12 \mul^2$, $x_{fp\pm}$ coalesce\footnote{Two signs of the angular momentum correspond to two opposite rotations. From now onward we will focus on the one with positive sign.}. This is the critical value of the angular momentum, namely $L _c= 2\sqrt{3} ~\mul$, at which the ISCO forms. The critical energy $E_c$  and angular velocity $\Omega_c$ 
can be easily found. Below these critical values of the parameters, the fixed points disappear. All we have are the unstable plunging trajectories of the particle. This is therefore a point of bifurcation.\\

\noindent
It is well known that bifurcations in dynamical systems possess stuctural similarities with  thermodynamic phase transition (see \cite{Bose_2019} for a short review). For instance, they both describe sudden, qualitative shifts in  system's behavior when a control parameter crosses a threshold. Pushing this analogy further, we  anticipate that  the energy, angular momentum, angular velocity and certain response functions  associated with  the fixed points  near  the ISCO exhibit  power law scaling relations akin to those found in continuous  phase transitions. Indeed it is easy to show, and we will  discuss later in more generality, the occurrence of such 
 scaling relations. Further, we find that  the relevant set of scaling exponents are same 
as that of the mean field exponents of  the van der Waals fluid. For example,   
at $E = E_c$
the angular momentum and the angular velocity satisfy a relation  of the form
\begin{equation}
L - L_c \sim (\Omega - \Omega_c)^\delta, ~~{\rm with}~\delta = 3.\ \nonumber
\end{equation}
\noindent
 These scaling behaviors assume additional significance for black holes in the anti-de Sitter (AdS) spacetime. This is because of the AdS/CFT correspondence.  AdS  differs from  the flat spacetime by an inherent length scale $(l)$,  arising from the negative cosmological constant. In dimensions greater than three,
 it is well known that the Schwarzschild black hole in asymptotic  AdS spacetime admits periodic as well as plunging trajectories for the geodesics~\cite{Festuccia:2008zx,Berenstein:2020vlp}. Consider, for instance, a five dimensional black hole in AdS where the corresponding phase-plane equations are exactly solvable. We get two fixed points with one being a center, representing a stable circular orbit, and the other is a saddle. 
  As the mass of the black hole is taken to zero, holding other parameters fixed, the saddle moves closer to the singularity and finally disappears. What remains is the center at $x = (l L)^{-1/2}$ of the resulting global AdS geometry \footnote{Unlike in flat spacetime, the inherent length scale $l$ allows for a stable circular orbit in AdS.}. On the other hand, if we keep the mass fixed and  lower the angular momentum of the particle, two fixed points approach each other, coalescing at ISCO. For $L$ below this threshold, we are only left with the plunging trajectories. This is again a bifurcation point and, as before, various scaling behaviors  are seen to arise for the trajectories in the vicinity of the ISCO. \\
  
\noindent  
The AdS/CFT correspondence associates a theory of classical gravity on AdS with a strongly coupled gauge theory on its boundary. When the bulk contains a black hole of size much larger than the AdS scale $l$, the gauge theory is in its deconfined phase with  a temperature
 dictated by the size of the black hole. Consequently, the bulk  geodesics now receive additional meaning in terms of the gauge theory  variables. On one hand, the plunging trajectories  get related to  the thermalization process in the gauge theory~\cite{Balasubramanian:2011ur}. On the other, the  stable bulk orbits, which do not classically thermalize, represent Regge trajectories in the dual theory~\cite{Festuccia:2008zx,Karlsson:2021duj,Berenstein:2020vlp,Dodelson:2022eiz}. Of course, processes like radiation and tunneling may render these states  metastable eventually~\cite{Dodelson:2022eiz}. This was already noted in  ~\cite{Festuccia:2008zx,Berenstein:2020vlp}. Here, the metastable stables (orbits) were only found in black hole geometries, where the boundary gauge theory lives on $R \times S^{d-1}$, with exponentially small imaginary part of quasi-normal modes indicating long lifetime before they  thermalize. However, in the case of AdS black-branes with infinite volume in the boundary theory $R \times R^{d-1}$, such states were shown to thermalize at O(1) times, indicating the absence of stable circular orbits. Binding energy of these classically stable orbits were computed in~\cite{Fitzpatrick:2014vua,Berenstein:2020vlp} in arbitrary dimensions. The bulk two-body problem discussed above is related to the double-twist operators in a holographic CFT spectrum~\cite{Dodelson:2022eiz,Kulaxizi:2018dxo}. Instead of four identical scalars, we choose a pair of heavy operators in the four point function. The heavy operators {$\Delta_H$} correspond to the black hole in the bulk dual around which the light operators {$\Delta_L$} probe the corresponding geometry. In the limit of large angular momentum, 
 the binding energy, when expressed in terms of relevant gauge theory quantities, expectedly, is found to match with
 the anomalous dimension of certain double twist operator of the theory, which are universal in all CFT's. Naturally, therefore one expects the correlations involving these double twist operators to carry  signatures of the bulk scaling behavior near the bifurcations. \\
 
 \noindent
We should remember that the double-twist states in the CFT spectrum alluded to above, are obtained by lightcone bootstrap\footnote{Even though the lightcone bootstrap formalism is valid for any CFT at large spin,  here we focus on holographic CFT's where $c_T \to \infty$.}, and they correspond to a discrete spectrum and are universal. When we include heavy states in the spectrum, the discrete states are broadened into resonances, which are  non-perturbative in spin. These can be identified with long-lived metastable states, or resonances in the dual CFT. Perturbatively in spin, we obtain the discrete spectrum of states that correspond to a stable orbit around a black hole. We study the heavy-light four point function in the $s$-channel where the double twist operators are exchanged. The narrow resonances in the heavy-light OPE can be identified with quasi-normal modes with a small imaginary part in the thermal two-point function. This is discussed and reviewed in a later section of the text. The eigenstate thermalization hypothesis~\cite{Srednicki_1999,D_Alessio_2016,lashkari2016eigenstatethermalizationhypothesisconformal,10.21468/SciPostPhys.9.3.034} is used as a bridge between the heavy-light OPE and the quasi-normal mode picture. When the decay due to radiation and tunneling can be neglected, the discrete spectrum of states is obtained, as usual in lightcone bootstrap analysis\footnote{This was studied in section 3 of~\cite{Dodelson:2022eiz}.}. The lightcone bootstrap involves correction to  Mean-field theory (MFT) OPE coefficients in the large spin expansion. This picture tells us that the orbit states, although to leading order are linked to large spin expansion of the heavy-light correlator in the CFT, there must also be sub-leading corrections to account for the gravitational radiation (the loss of energy due to radiation and the correction to the geodesic equations themselves) \footnote{The tunneling occurs as a non-perturbative correction to the binding energy~\cite{Festuccia:2008zx}.}. In order to make explicit the sub-leading corrections to MFT behavior, we perform a $\frac{1}{\Delta_H}$ expansion of the OPE coefficients and compute the corresponding anomalous dimensions. Thus, we are looking at a large spin and large dimension expansion of the OPE. We thus obtain the sub-leading corrections in $\frac{1}{\Delta_H}$ to the anomalous dimensions in the boundary CFT, which should eventually be matched to the corrections of the bulk orbits.
\\

\noindent 
The outline of the rest of the paper is as follows. In section-{\ref{two}}, we first develop  a phase-plane analysis of the  geodesics in stationary, spherically symmetric black hole background in arbitrary dimensions. It turns out that much about the nature of the trajectories can be learned without explicitly specifying the form of the metric.  In particular, it can be argued that if the phase portrait admits a center that exits only until a parameter reaches a threshold value, there must exist a saddle. At ISCO, the saddle 
 coalesces with the center, causing a bifurcation. Manifestations of this phenomenon are then seen through various scaling relations which we elaborate in the subsection-(\ref{two.1}). Subsection-{\ref{two.2}} includes explicit examples of bifurcations and critical behavior  in Schwarzschild black holes in flat as well as  in AdS spacetime. Further, we extend the analysis to  include Reissner-Nordstrom black holes in asymptotically flat spacetime. In section-(\ref{three}), we focus on the asymptotically AdS spacetimes in five spacetime dimensions  and explore the consequences of the critical behavior near the ISCOs in the dual CFT. To this end, in subsection (\ref{three.1}), we first extract the behavior of anomalous dimensions $\gamma$ of certain double twist operators from the contrasting behavior of center and the saddle, through AdS/CFT. Here, we point out the subtelities in interpreting $\gamma$ for the center as well as the saddle and also discuss the non-analyticity of $\gamma$ that appears in the limiting case, where the center and saddle coalesce to produce the ISCO. In the following subsection-(\ref{three.3}), we consider the bulk orbits from the perspective of the holographic CFT. We review the heavy-heavy-light-light four point bootstrap in the usual mean-field theory limit, matching the results from lightcone bootstrap.  In subsection-(\ref{three.4}), we discuss the corrections to mean-field theory behavior by expanding to next order in $\frac{1}{\Delta_H}$. We compute the effect of these corrections to the anomalous dimensions of the double-twist operators in the CFT (orbit states in the bulk). We comment on the possible-bulk interpretation of these results. We conclude in section-(\ref{four}) by summarizing our results and discussing possible future directions.



\section{Fixed points, ISCOs and Criticality} \label{two}

\noindent  In this section, we  develop  a phase-plane analysis of the  geodesics in stationary, spherically symmetric black hole background. In the first two subsections, we
look for ISCO, bifurcation and various scaling behaviors in the vicinity of ISCO without making reference to any particular black hole. The final subsection-(\ref{two.2}) includes
explicit illustrations of the above phenomena for Schwarzschild black hole in asymptotically flat and AdS spactimes and also for the Reissner-Nordstrom black holes in  flat spacetime.

\subsection{Fixed points}
We start with a general black hole metric in $d+1$ dimension
\begin{equation}
ds^2 = - f(r) dt^2 + \frac{dr^2}{f(r) }+ r^2 d\Omega^2_{d-1},\label{eq:hole}
\end{equation} 
where $d\Omega^2_{d-1}$ is the metric on unit $S^{d-1}$. At this stage we keep
$f(r)$ unspecified. Horizons appears where  the function $f(r)$ vanishes. We will indicate their locations as $r_h$. The trajectory of a massive particle in this geometry satisfies
\begin{equation}
- f \dot t^2 + f^{-1} \dot r^2 + r^2 \dot \phi^2 = -1.
\label{eq:one}
\end{equation}
The dots represent derivatives with respect to the proper time. In writing down the equation, we have made use of the  spherical symmetry of
the metric and focused on its equatorial plane. Here $\phi$ is the azimuthal coordinate. The
time translational and the rotational symmetries of the geometry provide us with two conserved quantities  for the particle-- the energy $E$ and
the angular momentum $L$. These are given by
\begin{equation}
E = f(r) \dot t ~~{\rm and} ~ L = r^2 \dot \phi.
\end{equation}
Using these conserved quantities, equation (\ref{eq:one}) can be recast into the form
\begin{equation} \label{Veff}
\dot r^2 = E^2 - V_{\rm{eff}}(r).
\end{equation}
We have introduced the potential
\begin{equation}
V_{\rm{eff}}(r) = f (r)\Big(1+ \frac{L^2}{r^2}\Big).
\label{eq:poten}
\end{equation}
\noindent
For our purpose,  it is instructive to rewrite  the differential equation in $r,\phi$ variables.  Using $dr/d\phi = \dot r/\dot\phi$,
we get 
\begin{equation}
\Big(\frac{dr}{d\phi}\Big)^2 = \frac{E^2}{L^2} r^4 - \frac{r^4}{L^2} f \Big(1 + \frac{L^2}{r^2}\Big).\nonumber
\end{equation}
Written in  $x = 1/r$, the equation simplifies 
\begin{equation}
\Big(\frac{dx}{d\phi}\Big)^2 = \frac{E^2}{L^2} - (1 + L^2 x^2) \frac{f(x)}{L^2}.
\label{eq:three}
\end{equation}
One further derivative reduces it to
\begin{equation}
\frac{d^2x}{d\phi^2} = -\frac{1}{2L^2} (1 + x^2 L^2) \frac{df}{dx} - x f.
\end{equation}
Defining $dx/d\phi = z$ as the new variable, we convert the above into two first order differential equations 
\begin{eqnarray}
&&\frac{dx}{d\phi} = z,\nonumber\\
&&\frac{dz}{d\phi} = -\frac{1}{2L^2} (1 + x^2 L^2) \frac{df}{dx} - x f.
\label{eq:fixed}
\end{eqnarray}
These two equations encode the details of the particle motion and provide us with the phase portraits.

\bigskip

\noindent The trajectories are controlled by the fixed points. The fixed points are the solutions of
\begin{equation}
z = 0, ~~{\rm and}~~\frac{1}{2L^2} (1 + x^2 L^2) \frac{df}{dx}  + x f =0.
\label{eq:six}
\end{equation}
Depending on the explicit form of $f(x)$, we could have one or multiple solutions.
As for now,  we denote  all solutions together as 
$x = x_{fp}$. From the first of these equations and from (\ref{eq:three}), we get the impact parameter $b = E/L$ for  the fixed point(s):
\begin{equation}
b^2 = \frac{E^2}{L^2} = (1 + L^2 x_{fp}^2)\frac{f(x_{fp})}{L^2}.
\label{eq:seven}
\end{equation}
Finally, solving the second equation of (\ref{eq:six}) along with (\ref{eq:seven}), we get
\begin{equation}
E =  \sqrt{  \frac{2 f(x_{fp})^2}{2 f(x_{fp}) + x_{fp} f^\prime(x_{fp})} },~~L  = \pm \sqrt{
- \frac{f^\prime(x_{fp})} {x_{fp} (2 f (x_{fp}) + x_{fp} f^\prime(x_{fp}))} }.
\label{eq:eight}
\end{equation}
Two signs of $L$ represent motions in two opposite directions. 
These equations are to be understood as follows. We assume the form of $f$ is such that 
the second equation of (\ref{eq:six}) offers one or more real positive solution for $x$.
Since a conservative system (in $\phi$!) can not have  attractive fixed points, 
fixed points could be either all centers, or all saddles or their combinations. We denoted those solutions
together as $x_{fp}$. For the energy associated with the trajectory $x = x_{fp}$,  to be real, the denominator in (\ref{eq:eight}) should satisfy $2 f (x_{fp}) + x_{fp} f^\prime(x_{fp}) > 0$. Then 
from the second equation, we see the angular momentum is real if $f^\prime(x_{fp}) < 0$. Note that the  angular velocity with the trajectory is given by
\begin{equation}
\Omega = \frac{d\phi}{dt} = \frac{x_{fp}^2 L f(x_{fp})}{E} =  \sqrt{ - \frac{f^\prime(x_{fp}) x_{fp}^3}{2}}.
\label{eq:nine}
\end{equation}
\noindent
We will now assume that $f(x)$ is such that it provides at least one  center.
Examples of this kind are in plenty. Take the global AdS metric with $f(x) = 1 + 1/(l^2 x^2)$. Substituting it in  (\ref{eq:six}), we see that
$x_{fp} = 1/\sqrt{lL}$. A linear stability analysis near this point suggests that it is a center providing us with a timelike classically stable circular orbit of radius $r_{fp} = 1/x_{fp}$. 
The energy associated with the orbit is $E = 1 + L/l$ and it follows from the first equation of (\ref{eq:eight}).
We thus have a center in AdS for any value of the angular momentum.  As $x_{fp}$ increases, not surprisingly, both $L, E$  decrease.  Consider now a four dimensional
Schwarzschild black holes, with $f(x) = 1 -  2\mul x$. A center appears when $L^2$  crosses a critical
value. This can be seen again from (\ref{eq:six}). It gives a center at 
\begin{equation}
x_{fp-} = \frac{1}{6L\mul} \Big(L - \sqrt{L^2 -12 \mul^2}\Big).\nonumber
\end{equation}
Clearly, stable circular orbits exist for any value of $L^2 \ge 12 \mul^2$, with an ISCO occurring at the minimum of $L$. 
The location of it is $x_{fp} = 1/(6\mul)$. For AdS-Schwarzschild for $d >2$, an analogous  story repeats. 
We will come back to these specific geometries in later sections.\\

\noindent
If $f(x)$ admits a  center, as we have assumed previously, there could be  two possible scenarios -- first, regardless of
the value of the angular momentum, the center survives and the other, as we reduce the angular momentum,
below a critical value, it disappears. First possibility is realized in global AdS and the other in various black holes. 
If the later happens, the following topological arguments suggest that, regardless of the exact functional form of $f(x)$, there must exist a saddle point.\\

\noindent
The argument is very standard. Let us imagine a simple closed curve $C$ enclosing a fixed point. We make sure that the point does not sit on $C$.
We can then define an angle
\begin{equation}
\psi = {\rm tan}^{-1}\Big(\frac{z^\prime}{x^\prime}\Big),\nonumber
\end{equation}
where $\prime$ denotes derivative with respect to $\phi$. As we move counterclockwise on $C$, $\psi$ changes
and we can associate an index to this closed curve as
\begin{equation}
I_C = \frac{1}{2\pi}[\psi]_C.\nonumber
\end{equation}
Here $[\psi]_C$ is the total change of $\psi$ over one circuit. Since $I_C$ is an integer, under continuous 
deformations of $C$, this number does not change. Therefore, it makes sense to associate $I_C$ to
the fixed point itself. While one finds  $I_C = 1$ for a center, if  $C$ encloses a saddle point, it is $-1$.\footnote{For more details, we refer to
the book Nonlinear Dynamics and Chaos, Steven H. Strogatz, section 6.8.} Now suppose we assume that $f(x)$ is such that there is an ISCO.
It means that, as we reduce $L^2$ (and $E$), there is a threshold value below which the center disappears.
But since the topological charge has to change from $1$ to $0$, this can only happen if a saddle point comes from somewhere and coalesces with the center reducing the topological charge by one unit.

\subsection{ISCOs and criticality: general features} \label{two.1}
We now argue that the point at which the saddle merges with the center, it is a very special point. In the following we will call it a critical point.
We indicate $L, E, \Omega, x$ associated with the critical point with a subscript $c$. The critical angular momentum, $L_c$, can be computed by minimizing
the second equation of (\ref{eq:eight}). 
\begin{equation}
\frac{\partial L^2}{\partial x_{fp}} = 0, ~~{\rm gives}~ f^{\prime\prime}(x_{fp}) = f^\prime (x_{fp})\Big(\frac{1}{x_{fp}} + \frac{2 f^\prime(x_{fp})}{f(x_{fp})}\Big).
\label{eq:ten}
\end{equation}
The solution of the last equation determines $x_c$, and, when substituted  it back to $L$, we get  $L_c$. Similarly, we can compute
$E_c, \Omega_c$. Having obtained the critical quantities, we now argue that, near the critical point, $L = L (E, \Omega)$ shows
scaling properties. To this end, we expand, $L(E_c, \Omega)$ near $\Omega_c$ as
\begin{eqnarray} 
L (E_c, \Omega) = L_c &+& (\Omega - \Omega_c)\Big(\frac{\partial L(E_c, \Omega)}{\partial \Omega}\Big)_{\Omega = \Omega_c} + (\Omega - \Omega_c)^2 \Big(\frac{\partial^2 L(E_c, \Omega)}{\partial \Omega^2}\Big)_{\Omega = \Omega_c}\nonumber\\
 &+& (\Omega - \Omega_c)^3\Big(\frac{\partial^3 L(E_c, \Omega)}{\partial \Omega^3}\Big)_{\Omega = \Omega_c} 
+ ...\nonumber
\end{eqnarray}
To compute the partial derivatives, we use $L$ coming from (\ref{eq:seven}),
\begin{equation}
L(\Omega, E_c) = \sqrt{\frac{E_c^2 - f(x)}{x^2 f(x)} },
\label{eq:eleven}
\end{equation}
and  $\Omega$ from (\ref{eq:nine}). Writing 
\begin{equation}
\frac{\partial L}{\partial \Omega} = \frac{\partial L}{\partial x}\Big/ \frac{\partial \Omega}{\partial x},\nonumber
\end{equation}
we see that it vanishes at $\Omega = \Omega_c$ as ${\partial L}/{\partial x}$ is zero at $x = x_c$, while ${\partial \Omega}/{\partial x}$ is finite.  Further, $\partial^2 L/\partial x^2$ also vanishes at $x = x_c$. This can be seen as follows.  We
start with (\ref{eq:eleven}). We take  two derivatives of $L$ with respect to $x$ keeping  $E$ fixed to  $E_c$. To this, we substitute
$E_c$, following from the first equation of (\ref{eq:eight}) and the one for $f^{\prime\prime}(x_c)$ arising from (\ref{eq:ten}). A final simplification provides the desired result, that is
\begin{equation}
\Big(\frac{\partial^2L(E_c, \Omega)}{\partial \Omega^2}\Big)_{\Omega = \Omega_c} = 0.\nonumber
\end{equation}
Therefore, we conclude that near the critical point, with $E$ set to $E_c$
\begin{equation}
L - L_c \sim (\Omega - \Omega_c)^\delta, ~~{\rm with} ~\delta = 3.
\end{equation}
Similarly close to $E = E_c$ and $\Omega = \Omega_c$, we find
\begin{equation}
\Big(\frac{\partial L}{\partial \Omega}\Big)_{E, \Omega = \Omega_c} \sim (E - E_c).\nonumber
\end{equation}
Hence the response function, analogous to isothermal compressibility for fluid,
\begin{equation}
K_E = - \Big(\frac{1}{\Omega} \frac{\partial \Omega}{\partial L}\Big)_{ E, \Omega = \Omega_c} \sim (E - E_c)^{-\gamma}, ~~{\rm with} ~\gamma = 1.
\end{equation}
To find out the other scaling properties, we require explicit form of $f(x)$. We will return to it in the following section. As we have argued previously, at criticality, we have
\begin{equation} 
\frac{\partial L}{\partial \Omega}= 0, \qquad \frac{\partial^2 L}{\partial \Omega^2} = 0.
\label{eq:critone}
\end{equation}
We can immediately draw an analogy between these equations and the equations that identify the
critical point of a second order thermodynamic phase transition. For illustrative  purpose, take van der Waals fluid as an example. At the critical point,
the pressure $P$ and the volume $V$ satisfy
\begin{equation}
\frac{\partial P}{\partial V}= 0, \qquad \frac{\partial^2 P}{\partial V^2} = 0.
\label{eq:vdw}
\end{equation}
The similarities of the equations (\ref{eq:critone}) and (\ref{eq:vdw}) beg for a correspondence between pressure and angular momentum on one
hand and volume and angular velocity on the other. Proceeding further, from (\ref{eq:critone}), it follows
\begin{equation} 
\frac{\partial (L^2)}{\partial \Omega} = 0, \qquad \frac{\partial^2(L^2)}{\partial \Omega^2} = 0.
\label{eq:oneaa}
\end{equation}
We can use these two equations to identify the critical point.
First,  from (\ref{eq:three})  we have,
\begin{equation}
L^2 = \frac{E^2 - f(x_{fp})}{x_{fp}^2 f(x_{fp})}.
\end{equation}
Now using (\ref{eq:nine}), we write
\begin{equation}
\frac{\partial (L^2)}{\partial \Omega}=  \frac{\partial (L^2)}{\partial x_{fp}} /\frac{\partial (x_{fp})}{\partial \Omega}  
=- \frac{ 2 E^2 f(x_{fp}) - 2 f^2(x_{fp}) + E^2 x_{fp} f^\prime(x_{fp}) } {x_{fp}^3 f^2(x_{fp})} = 0.
\end{equation}
Therefore when $x_{fp} = x_c$,
\begin{equation}
 2 E^2 f(x_{c}) - 2 f^2(x_{c}) + E^2 x_{fp} f^\prime(x_{c}) = 0.
 \label{twoaa}
 \end{equation}
 The second equation in (\ref{eq:oneaa})  gives\footnote{To get to this one needs to use $f^\prime(x_{fp})$ from 
 the  equation (\ref{twoaa}).}
 \begin{equation}
 14 E^2 f^2(x_{c}) - 8 f^3(x_{c}) + E^4 \Big(x_{c}^2 f^{\prime\prime} (x_{c}) - 6 f(x_{c})\Big) = 0.
 \label{twobb}
 \end{equation}
Solving the above two equations, we determine $E_c$ and $x_c$. For example, for Schwarzschild black hole, $f(x) = 1- 2 \mul x$ and we get, 
\begin{equation} 
E_c = \frac{2 \sqrt 2}{3}, \qquad x_c = \frac{1}{6\mul}.
\end{equation}
Further rewriting
\begin{equation}
L^2 = \frac{E^2 - f(x_{fp})}{x_{fp}^2 f(x_{fp})} = - \frac{E^2 x_{c} f^\prime(x_{c})}{2  x_{c}^2 f^2(x_{c})},
\end{equation}
we obtain
\begin{equation}
\frac{L_c \Omega_c}{E_c} = -\frac{x_c f^\prime(x_c)}{2 f(x_c)}.
\end{equation}
With $x = 1/(6 m)$, for Schwarzschild, we arrive at 
\begin{equation}
\frac{L_c \Omega_c}{E_c}= \frac{1}{4}.
\end{equation}
This is analogous to the relation $P_c V_c/T_c = 3/8$ satisfied by the van der Waals fluid at the second order phase transition point.

\subsection{ISCOs and criticality: explicit illustrations} \label{two.2}

\noindent In the following, we provide some explicit examples.\\

\bigskip

\begin{figure}[h!]
\centering
\includegraphics[width=6in]{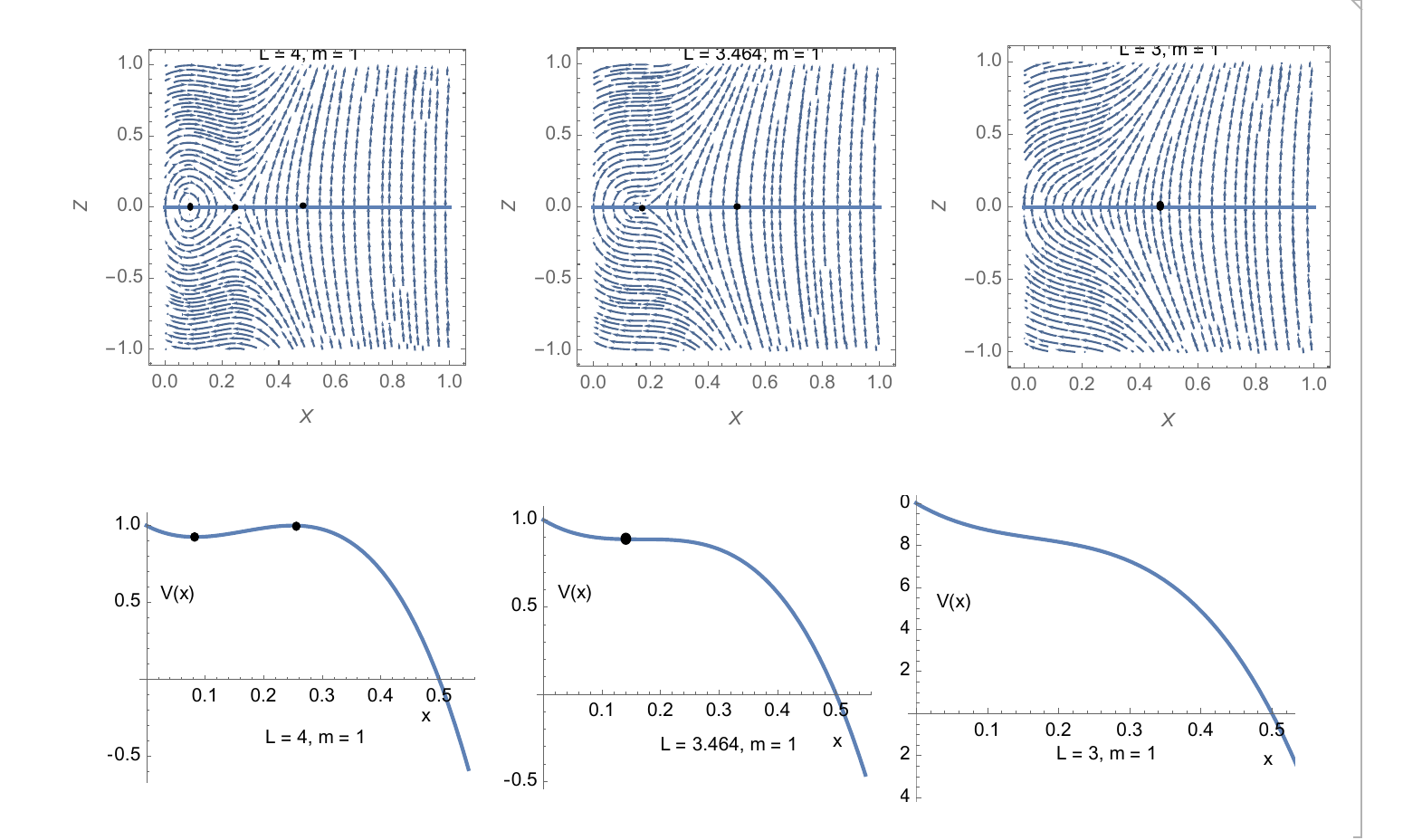}
\caption{The phase portrait of four dimensional Schwarzschild black hole is shown in the first row for fixed $m$
and $L$ decreasing from left to right. The right black dots common to all the three plates represent the location of the horizon. Other dots are placed to indicate the center and the saddle come close and coalesce and disappear. 
The second row shows the same in potential $V(x)$ introduced in (\ref{eq:poten}).}
\label{fi:zerozero}
\end{figure}

\noindent{\bf Schwarzschild}: We start with the Schwarzschild geometry in four dimensions. Phase plane analysis of the system was carried out earlier, see~\cite{Dean:1999edw,Jia:2017nen}.  We focus on the ISCO, bifurcation and corresponding critical behavior. Here, the 
saddle and center appear at
\begin{equation}
x_{fp-} = \frac{1}{6L\mul}\Big(L - \sqrt{L^2 - 12 \mul^2}\Big), ~~x_{fp+} = \frac{1}{6L\mul}\Big(L + \sqrt{L^2 - 12 \mul^2}\Big),
\label{eq:xfp}
\end{equation}
respectively. 
The locations of the fixed points are shown in figure (\ref{fi:zerozero}).
Explicit dependence on $m$ in these expressions and the rest  can be avoided 
with the use of  scaled variables $\tilde x = \mul x, \tilde L = L/\mul,  \tilde E = E$  and $\tilde \Omega = \mul \Omega$. 
In terms of these new variables, the above equations reduce to
\begin{equation}
\tilde x_{fp-} = \frac{1}{6\tilde L}\Big(\tilde L - \sqrt{L^2 - 12}\Big), ~~\tilde x_{fp+} = \frac{1}{6\tilde L}\Big(\tilde L + \sqrt{\tilde L^2 - 12 }\Big).\nonumber
\end{equation}
From now on
we will use these scaled variables and,  to avoid the cluttering, we will not use the tildes explicitly. The scaled energy, angular momentum and the angular velocities now take the forms
\begin{equation}
E = \frac{1 - 2 x_{fp}}{\sqrt{1-3x_{fp}}}, ~~L = \frac{1}{\sqrt{x_{fp}(1-3 x_{fp})}}, ~~\Omega = x_{fp}^{\frac{3}{2}}.
\label{eq:scaled}
\end{equation}
Further,  with (\ref{eq:scaled}),   (\ref{eq:seven}) can be rearranged to write
\begin{equation}
L = \sqrt{\frac{1 - E^2 - 2 \Omega^{\frac{2}{3}}}{2 \Omega^2 -  \Omega^{\frac{4}{3}}} }.
\label{eq:eos}
\end{equation}
We note that the two fixed points in (\ref{eq:xfp}) coincide for  $L_c^2 = 12 $ at $x_{c} = 1/6$, producing an ISCO. Using scaled variables, $\bar E = E/E_c, \bar L = L/L_c, \bar \Omega = \Omega/\Omega_c$, we can write (\ref{eq:eos}) as
\begin{equation}
\bar L(\bar E, \bar \Omega) = \sqrt{\frac{9 - 8 \bar E^2 - 3 \bar \Omega^{\frac{2}{3}}}{\bar \Omega^2 - 3\bar\Omega^{\frac{4}{3}}} }.
\label{eq:eosone}
\end{equation}
Considering the  similarities between the present system and the van der Waals fluid, we may consider (\ref{eq:eos}) or (\ref{eq:eosone}) as the equation of state. 
\begin{figure}[h!]
\centering
\includegraphics[width=4in]{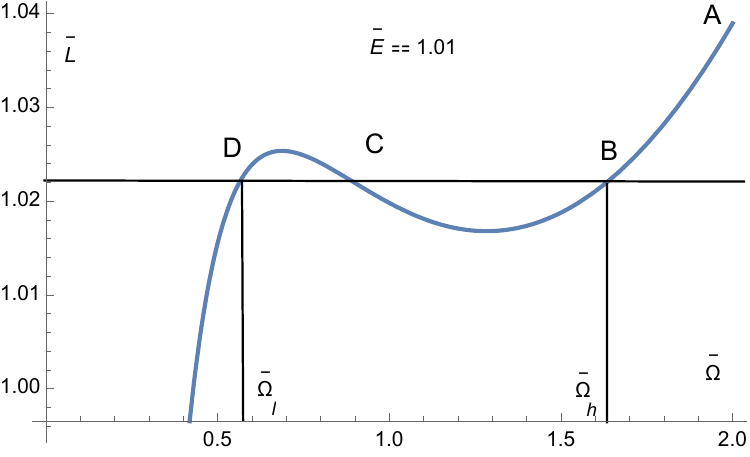}
\caption{The behavior of the angular momentum $\bar L$ as a function of $\bar \Omega$ for fixed $\bar E$ in the vicinity of the critical 
point. Note, in terms of the barred variables, the critical point is at $(\bar L_c, \bar \Omega_c, \bar E_c) = (1,1,1)$.}
\label{fi:zero}
\end{figure}
\noindent
As the exponents  $(\delta, \gamma)$ have been computed earlier, we now find $\beta$ here. We proceed with the dependence of $\bar L$ on $\bar\Omega$  as shown in
figure (\ref{fi:zero}). As we reduce $\bar L$ from a large value, the system moves down following the curve
$AB$ and then, via an analogue of Maxwell's construction, jumps to the point D. This is a point on the coexistence
curve, where given fixed angular momentum and energy, two states with different angular velocity ($\bar \Omega_h, \bar\Omega_l$)  may coexist. Therefore the following equation must hold:
\begin{equation}
\sqrt{\frac{9 - 8 \bar E^2 - 3 \bar \Omega_h^{\frac{2}{3}}}{\bar \Omega_h^2 - 3\bar\Omega_h^{\frac{4}{3}}} }
= \sqrt{\frac{9 - 8 \bar E^2 - 3 \bar \Omega_l^{\frac{2}{3}}}{\bar \Omega_l^2 - 3\bar\Omega_l^{\frac{4}{3}}} }.
\end{equation}
Solving for $\bar E$, we find
\begin{equation}
\bar E = \frac{1}{2}{\sqrt{\frac{3}{2}}} \sqrt{ 3 - \bar\Omega_h^{\frac{2}{3}} + \frac{ \bar \Omega_h^{\frac{4}{3}}}
{\bar\Omega_h^{\frac{2}{3}} + \bar\Omega_l^{\frac{2}{3}} + \frac{\bar\Omega_l^{\frac{4}{3}}}{-3+  \bar\Omega_h^{\frac{2}{3}}} } }.
\end{equation}
When $\bar E$ is near its critical value, we can expand $\bar \Omega_{h/l}  = 1 \pm \epsilon$ for small
$\epsilon$. Substituting these in the above equation, to the second order in $\epsilon$, we  get
\begin{equation}
 (\bar E - \bar E_c) \sim \epsilon^2
\sim (\bar \Omega_h - \bar \Omega_l)^2. 
\label{eq:beta}
\end{equation}
Therefore we get  $(\bar \Omega_h - \bar \Omega_l) \sim  (E - E_c)^\beta$,
 with $\beta = 1/2$.\\
 
\noindent
We close the discussion on  the Schwarzschild black hole with an outline on how to find the complete coexistence curve. Our analysis will follow  the method proposed in~\cite{Cui:2025bfr}. Given the angular momentum (\ref{eq:eosone}),
the analogue of the Gibbs free energy (at constant $\bar E$) function can be found as
\begin{equation}
d\bar  G = \bar \Omega d\bar L = \bar L \bar \Omega  - \bar L d\bar\Omega.
\end{equation}
Integrating on the coexistence curve,
\begin{equation}
\bar G (\bar\Omega_h) - \bar G(\bar\Omega_l)
=  \sqrt{ \frac{9 - 8 \bar E^2 - 3 \bar \Omega_h^{\frac{2}{3}}} {\bar \Omega_h^2 - 3 \bar\Omega_h^{\frac{4}{3}} } }
(\bar \Omega_h - \bar\Omega_l) - \int_{\bar \Omega_l}^{\bar \Omega_h} \bar L d\bar\Omega.
\end{equation}
Simplifying and setting the resulting expression to zero, we obtain
\begin{eqnarray}
&&-3 \sqrt{8 \bar E^2 -9} ~\Big({\rm EllipticE}\Big[{\rm sin}^{-1}(\bar \Omega_h^{1/3}/{\sqrt{3}}), -\frac{9}{8\bar E^2 -9}\Big]\nonumber\\
&&\qquad- {\rm EllipticE}\Big[{\rm sin}^{-1}(\bar \Omega_l^{1/3}/{\sqrt{3}}), -\frac{9}{8\bar E^2 -9}\Big] \Big)\nonumber\\
&&\qquad\qquad+ 
\sqrt{ \frac{9 - 8 \bar E^2 - 3 \bar \Omega_h^{\frac{2}{3}}} {\bar \Omega_h^2 - 3 \bar\Omega_h^{\frac{4}{3}} } }
(\bar \Omega_h - \bar\Omega_l) = 0.
\label{fivedd}
\end{eqnarray}
\noindent
Since $\bar E, \bar L$ remain same on the curve, following~\cite{Cui:2025bfr}, we use  the ansatz $\bar\Omega_l = \phi(\bar\Omega_h)/{\bar\Omega_h}, \bar\Omega_h =\phi(\bar\Omega_l)/{\bar\Omega_l}$,
to get ,
\begin{equation}
\bar E = \frac{1}{2\sqrt 2} \sqrt{(\bar L^2 + \frac{3}{\bar\Omega_h^{4/3}})(3 \bar\Omega_h^{4/3} - \bar\Omega_h^2)}
= \frac{1}{2\sqrt 2} \sqrt{(\bar L^2 + \frac{3}{\bar\Omega_l^{4/3}})(3 \bar\Omega_l^{4/3} - \bar\Omega_l^2)}.
\end{equation}
Solving the above, we get
\begin{equation}
\bar L =\sqrt{ \frac{{3 \phi^{2/3} \bar\Omega_h^{4/3} - 3\bar\Omega_h^{8/3}}}{-\phi^2 + 3  \phi^{4/3} \bar\Omega_h^{2/3}
- 3 \bar \Omega_h^{10/3} + \bar\Omega_h^4}} .
\end{equation}
Substituting $\bar L$ in the energy expression,
\begin{equation}
\bar E = \frac{1}{2\sqrt 2} \sqrt{3  - \bar\Omega_h^{2/3}}
\sqrt{ \frac{  3( \phi^2 - 3 \phi^{4/3} \bar\Omega_h^{2/3} + 3 \bar\Omega_h^{10/3} - \phi^{2/3} \bar\Omega_h^{8/3})}
{ \phi^2 - 3 \phi^{4/3} \bar\Omega_h^{2/3} + 3 \bar\Omega_h^{10/3} - \bar\Omega_h^4 }}.
\end{equation}
We now use the above in (\ref{fivedd}) along with $\bar\Omega_l = \phi(\bar\Omega_h)/\bar\Omega_h$ to
have an equation relating $\phi$ and $\bar\Omega_h$. Choosing $\bar\Omega_h$ and solving for $\phi$
leads us to the coexistence curve.\\

\bigskip

\noindent{\bf AdS-Schwarzschild}: To keep the computations tractable, we discuss here the five dimensional AdS-Schwarzschild black hole.  The pattern however is similar as long as  $d > 2$.  Exact solutions of (\ref{Veff}) in four dimensions can be found in~\cite{Cruz:1994ir}. Our interest will be on the phase portrait, bifurcation and corresponding critical behavior. The relevant  function  is $f(x) = 
1 + 1/(l^2 x^2) - 2 \mul x^2$. 
\begin{figure}[h!]
\centering
\includegraphics[width=5in]{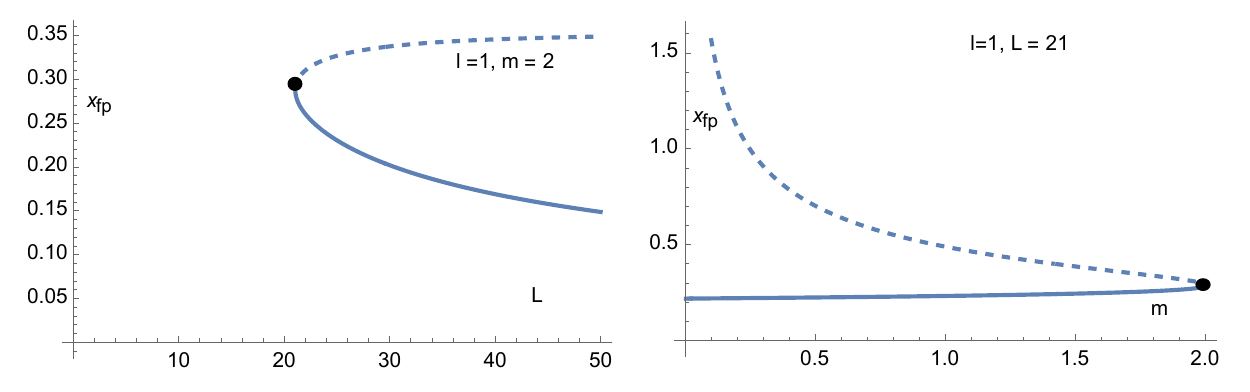}
\caption{The real positive solutions of the second equation in (\ref{eq:x}) as a function of $L$ (left) and $m$ (right). While the dashed line represents the saddle, the solid line is for the center.  In both the figures, the solid dots are the locations ISCO. }
\label{fi:oneone}
\end{figure}
The fixed points follow from (\ref{eq:six})
\begin{equation}
z = 0, ~~4 l^2 L^2 \mul  x^6 + 2 l^2 \mul x^4 - l^2 L^2 x^4 +1 = 0.
\label{eq:x}
\end{equation}
\begin{figure}[h!]
\centering
\includegraphics[width=6in]{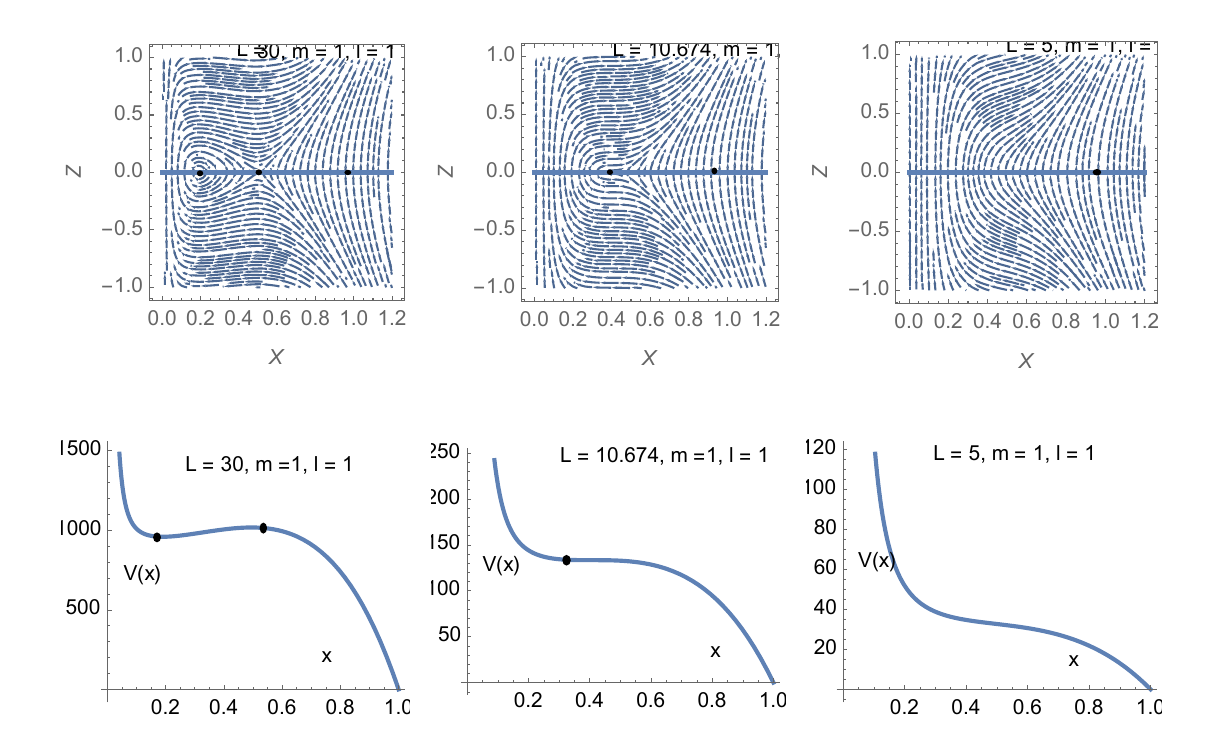}
\caption{First row shows the nature of fixed points of AdS-Schwarzschild black hole as we reduce angular
momentum $L$, keeping $m$ and $l$ fixed to $1$. We see the saddle and the center merging at 
$L = L_c = 10.67$ to form an ISCO. If we further lower $L$, no fixed point is seen. Note that 
the horizontal axis is $x = 1/r$. The location of the horizon is shown by a vertical green line. The second row 
is the plot of the effective potential $V(x)$ with $x$. While the minimum of the potential, shown with a dot, represents 
a center, the maximum corresponds to a saddle. We see that at $L = L_c$ maximum and minimum coalesce and then disappear.}
\label{fi:one}
\end{figure}
The second equation possesses  at most two real positive roots \footnote{One can check that $2\mul  - L^2$ is negative and one can use the rule of signs.} which we call together as $x_{fp}$. The smaller one
represents a center and the larger is a saddle, see figure (\ref{fi:oneone}). Also see figure (\ref{fi:one}) for the phase portrait and other details of the solution.  At a critical value of $L^2$, center and saddle merge to form an ISCO
and below this critical value of angular momentum, there are no fixed points.  Hence this is a point of bifurcation.\\

\begin{figure}[h!]
\centering
\includegraphics[width=3in]{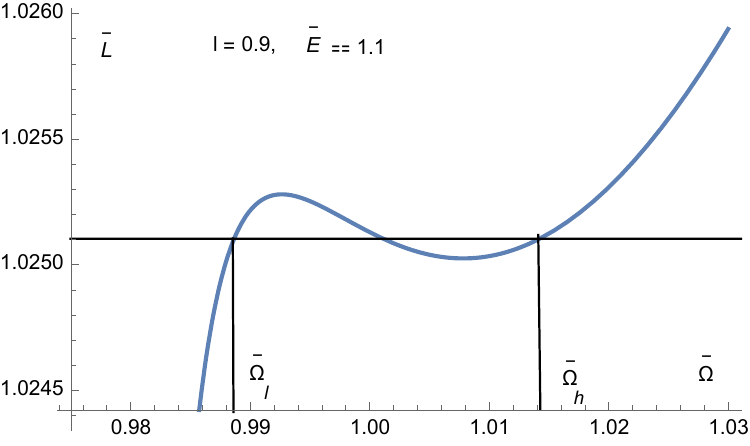}
\caption{Behavior of angular momentum as a function of angular velocity at fixed energy, $\bar E = 1.1$, just above the critical point for AdS Schwarzschild black hole. The cosmological constant has been set to  $l = 0.9$. }
\label{fi:five}
\end{figure}

\noindent
Before we find this critical value, we 
simplify various expressions using scaled\footnote{The rescaling to tilde variables is dimension dependent.} variables 
 $\tilde L = L/\sqrt{\mul}, \tilde \Omega = \Omega\sqrt{\mul}, \tilde l =  l/\sqrt{\mul}$  and $\tilde x = x\sqrt{\mul}$. 
 In these variables,  the second equation in (\ref{eq:x}) simplifies to
\begin{equation}
4 \tilde l^2 \tilde L^2   \tilde x^6 +( 2 \tilde l^2  - \tilde l^2 \tilde L^2) \tilde x^4 +1 = 0.
\label{eq:xx}
\end{equation}
 Unless otherwise mentioned, we will work with the scaled variables and, 
 for notational simplicity, we will avoid using the tildes explicitly. The angular momentum, energy and the angular velocity now have the following forms
\begin{equation}
L^2 = \frac{1 + 2 l^2 x_{fp}^4}{l^2 x_{fp}^4 (1 - 4 x_{fp}^2)}, ~~E^2 =\frac{ [l^2 x_{fp}^2 (1 - 2 x_{fp}^2) +1]^2}{l^4 x_{fp}^4 (1 - 4 x_{fp}^2)},
~~\Omega^2 = \frac{1}{l^2} + 2 x_{fp}^4.
\label{eq:leo}
\end{equation}
The equation of state follows from (\ref{eq:seven}) and can be written as 
\begin{equation}
L^2 = \frac{2 l (-2 \sqrt{2} + \sqrt{2} l^2 \Omega^2 + (E^2 -1) l\sqrt{l^2 \Omega^2 -1} )}
{ \sqrt{l^2 \Omega^2 -1} (4 - 2 l^2 \Omega^2 + l\sqrt{2(l^2 \Omega^2 -1)})}.
\end{equation}
\noindent
Differentiating $L^2$ with respect to $x_{fp}$ and setting it to zero, we get the location as well as the critical value
of angular momentum and other associated quantities.
\begin{eqnarray}
&&x_c^2 = {\frac {A^{\frac{1}{3}}}{2 l^2} - \frac{1}{A^{\frac{1}{3}}} } ,\nonumber\\
&& L_c^2  =   {\frac{2 (2 l^4  A - l^2  A^{\frac{5}{3}}  - 4 l^6  A^{\frac{1}{3}} )}
{(2 A^{\frac{2}{3}} -  l^2 A^{\frac{1}{3}}  - 4 l^2 )(A^{\frac{2}{3}} - 2l^2 )^2} },\nonumber\\
&&E_c^2 = \frac{(A^{\frac{4}{3}} - 6 l^2 A^{\frac{2}{3}} - l^2 A  + 4 l^4+ 2 l^4 A^{\frac{1}{3}})^2}
{ l^2 A^{\frac{1}{3}} (A^{\frac{2}{3}} - 2 l^2)^2 ( l^2 A^{\frac{1}{3}} + 4 l^2 - 2 A^{\frac{2}{3}})},\nonumber\\
&&\Omega_c^2 = \frac{A^{\frac{2}{3}}}{2 l^4} + \frac{2}{A^{\frac{2}{3}}} - \frac{1}{l^2}.
\label{eq:crit}
\end{eqnarray}
where $A = l^4  (1 + \sqrt{1 + 8/l^2})$. Regardless of the value of $l$, it can be checked that
all the expressions on the right hand side of all the equations are positive.
It turns out to be convenient to work with a set of new variables $\bar L = L/L_c, \bar E = E/E_c, \bar \Omega = \Omega/\Omega_c$, such that the critical point occurs at $(1,1,1)$.
In the vicinity of  the critical point, for fixed $l$, behavior of the angular momentum as a function of $\bar\Omega$ is shown in figure (\ref{fi:five}). \\

\noindent
The computation of $\beta$ is similar to that of the previous example. However, because of the complicated equation of state,  calculations are somewhat messy. Nonetheless, we find the value remains unchanged and we get $\beta = 1/2$.\\

\bigskip

\noindent{\bf Reissner-Nordstr\"om}:  Here $f(x) = 1 - 2 m x + q^2 x^2$, where the parameters $m, q$ are related to
the mass and charge of the black hole respectively \footnote{Motion of a test particle in Reissner-Nordstr\"om background was considered previously, see for example \cite{Pugliese:2010ps}.}. The horizons exists for $m  \ge q$ and their locations are 
\begin{equation}
x_{h\pm} = \frac{1}{q^2} \Big(m \pm \sqrt{m^2 - q^2}\Big).
\label{eq:horizon}
\end{equation}
The fixed points follow from  the zeros of equations in (\ref{eq:fixed}). These are $z = 0, x = x_{fp}$ where $x_{fp}$ denotes the  real positive
solutions of the cubic equation
\begin{equation}
2 L^2 q^2 x^3 - 3 L^2 m x^2 + (L^2 + q^2)x - m = 0.
\label{eq:fp}
\end{equation}
The expression on the left  is negative at $x = 0$ and diverges as $x^3$ for large positive $x$.
It therefore must have crossed zero at least once or at most three times. Consequently, there could be at least one or at most 
three \footnote{Since the coefficients are always positive, 3 positive real roots also  follow from Descartes rule of signs.} fixed points.\\

\noindent As previously done, we work with the scaled variables $\tilde x = mx, \tilde q = q/m, \tilde L = L/m,
 \tilde E = E$ and $\tilde\Omega = m \Omega$ for rest of the analysis. Equations \ref{eq:horizon} and (\ref{eq:fp}) now simplify to
 \begin{equation}
 x_{h\pm} = \frac{1}{q^2} \Big(1 \pm \sqrt{1 - q^2}\Big),
 \end{equation}
 and
\begin{equation}
 2 L^2 q^2 x^3 - 3 L^2 x^2 + (L^2 + q^2) x - 1 = 0,
 \label{eq:fp1}
 \end{equation}
 respectively. For brevity, we do not use tildes explicitly.
The angular momentum, energy and angular velocity associated to the fixed points are
\begin{equation}
L^2 = \frac{1 - q^2 x_{fp} } {2 q^2 x_{fp}^3 - 3  x_{fp}^2 +x_{fp}},~~E^2 = \frac{(q^2 x_{fp}^2 - 2  x_{fp} +1)^2}{2 q^2 x_{fp}^2 - 3 x_{fp} +1}, ~~\Omega^2 = (1 - q^2 x_{fp}) x_{fp}^3.
\label{eq:le}
\end{equation}
For these quantities to be real, we require
\begin{equation}
x_{fp} < \frac{1}{q^2}, ~~2 q^2 x_{fp}^2 - 3  x_{fp} + 1 > 0.\nonumber
\end{equation}
The later equation leads to the inequalities
\begin{equation}
x_{fp} < \frac{3}{4q^2} - \frac{1}{4q^2}\sqrt{9  - 8 q^2} = x_{fp-}
~~{\rm or},
~x_{fp} > \frac{3}{4q^2} + \frac{1}{4q^2}\sqrt{9  - 8 q^2} = x_{fp+}.\nonumber
\end{equation}
We additionally require the fixed points to lie outside the outer horizon of the black hole. Note  that with mass  now scaled to $1$, horizons exist only for the scaled charge  satisfying $0 \le q \le 1$.\\

\begin{figure}[h!]
\centering
\includegraphics[width=5in]{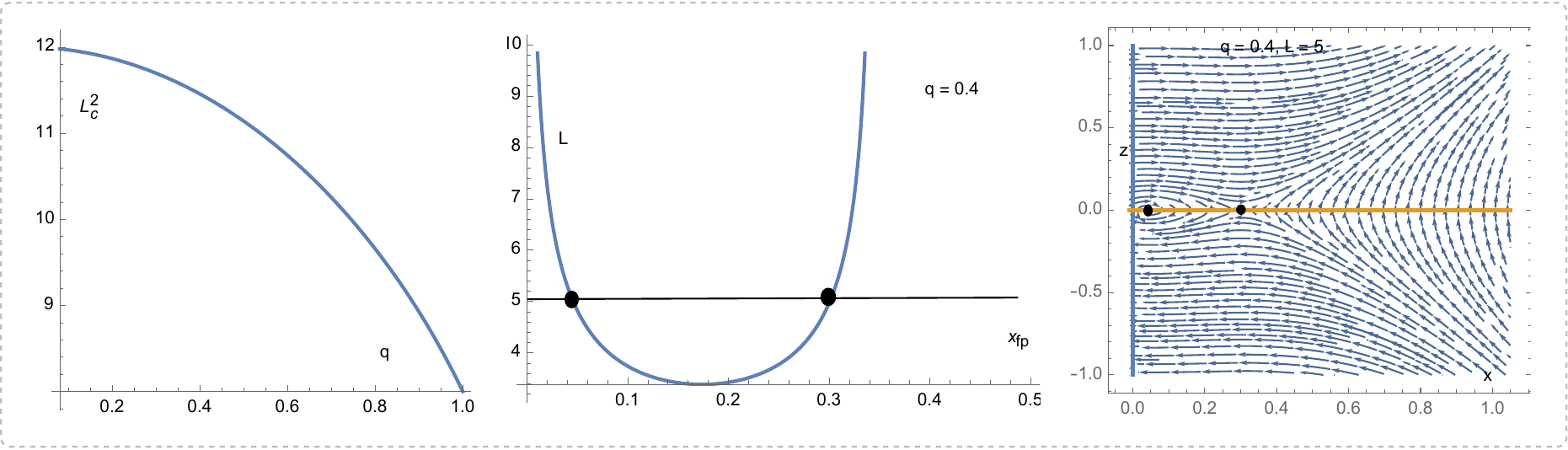}
\caption{The figure on the left shows the dependence of  $L_c^2$ on $q$. $L_c$ represents the value of angular momentum at which  the ISCO is formed. As we increase $L$ above $L_c$, the ISCO splits into two fixed points. An illustration of this phenomenon is shown  in the central figure for $q = 0.4$. A horizontal line
cuts the graph at two points. Corresponding values of $x_{fp}$ represent the locations of the two fixed points. 
As we reduce $L$, these two points approach each other coalescing at $L_c$. The phase portrait is drawn in the final figures.  It shows  that the one with smaller $x_{fp}$ corresponds to a center and, the larger one, a saddle. }
\label{fi:four}
\end{figure}

\noindent Solutions of the cubic equation (\ref{eq:fp1}) are easily found. These \footnote{We have written  them in complex form but they can be written in real form by using Viete's  trigonometric substitution formulae.} are
\begin{eqnarray}
&&x_{fp1} = \frac{1}{2q^2} - \frac{6 q^2 (q^2 + L^2 ) - 9 L^4}{2^{\frac{2}{3}}  q^2 L^2A^{\frac{1}{3}}} + \frac{A^{\frac{1}{3}}}{2^{\frac{1}{3}} q^2 L^2},\nonumber\\
&&x_{fp2} = \frac{1}{2q^2}  + \frac{ (1 + i \sqrt{3})(6 q^2 L^2 (q^2 + L^2) - 9 L^4)}{2^{\frac{2}{3}} 6 q^2 L^2 A^{\frac{1}{3}}}
- \frac{(1 - i \sqrt{3}) A^{\frac{1}{3}} }{2^{\frac{1}{3}} \times 12 L^2 q^2},\nonumber\\
&&x_{fp3} = \frac{1}{2q^2}  + \frac{ (1 - i \sqrt{3})(6 q^2 L^2 (q^2 + L^2) - 9 L^4)}{2^{\frac{2}{3}}\times 6 q^2 L^2 A^{\frac{1}{3}}}
- \frac{(1 + i \sqrt{3}) A^{\frac{1}{3}} }{2^{\frac{1}{3}} \times 12 L^2 q^2},
\end{eqnarray}
where 
\begin{eqnarray}
&&A = 54L^6 - 54 q^2 L^6 + 54 q^4 L^4 \nonumber\\
&& \qquad\qquad  + \sqrt{(54 L^6 - 54 q^2 L^6 + 54 q^4 L^4)^2 + 4( 6 q^2 L^2 (q^2 + L^2) - 9 L^4)}.\nonumber
\end{eqnarray}
It can be checked that  $x_{fp1}$  takes minimum value $1/q^2$ for $L = 0$ and increases monotonically with $L$. Therefore, this fixed point always stays behind the outer horizon and is not observed by  an asymptotic observer. 
On the other hand the rest two, namely $x_{fp2}$ and $x_{fp3}$  carrying opposite topological charges, arise only if 
$L$ becomes larger than  a critical value. Following our  pervious notational scheme, we call this value of $L$ to be  $L_c$.  The said nucleation point
is determined by setting the discriminant of (\ref{eq:fp1}) to zero, or else, using (\ref{eq:le}) along with equation (\ref{eq:oneaa}). The dependence of  $L_c$ on $q$, though  can be analytically obtained, is not particularly  illuminating. We therefore resort to a graphical representation of it in figure (\ref{fi:four}).\\

\begin{figure}[h!]
\centering
\includegraphics[width=3in]{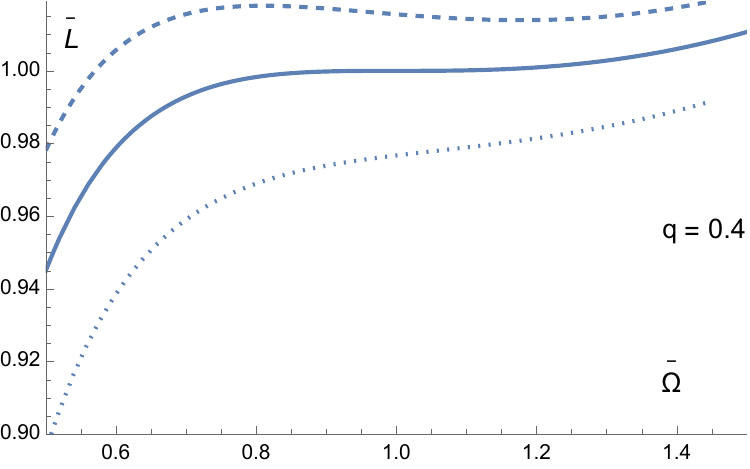}
\caption{This graph plots the angular momentum $\bar L= L/L_c$ as a function of $\bar\Omega = \Omega/\Omega_c$, arising from  (\ref{eq:lrn}), for a given charge. We fix the charge  $q$ to be $0.4$. Three plots are for three different values of energies restricted  in the vicinity of the critical point,  $E_c = .941, L_c = 3.385, \Omega_c = .071$.
While the solid curve  is for $E = E_c$,  the dashed and the dotted ones are for $E > E_c$ and $E < E_c$ respectively.}
\label{fi:lom}
\end{figure}

\noindent The equation of state, analogous to (\ref{eq:eos}), follows from setting (\ref{eq:three}) to zero and  solving for $L$. To express $L$ as a function of $E$ and $\Omega$, we  use the last equation of (\ref{eq:le})  to remove its dependence on  $x_{fp}$. The final expression is somewhat messy but useful for subsequent computations. We get
\begin{eqnarray}
L =\sqrt{ \frac{ E^2 + 2(A_1 - \frac{1}{2} B_1) - q^2(A_1 - \frac{1}{2}B_1)^2 -1}{  (A_1 - \frac{1}{2}B_1)^2
(1 - 2 (A_1 - \frac{1}{2} B_1) + q^2 (A_1 - \frac{1}{2} B_1)^2)}},
\label{eq:lrn}
\end{eqnarray}
where
\begin{eqnarray}
&&A_1 = \frac{1}{4q^2} + \frac{1}{2} \sqrt{\frac{1}{4q^4}  + \alpha + \frac{4\Omega^2}{3\alpha q^2}},\nonumber\\
&&B_1 = \sqrt{ \frac{1}{2q^4} - \alpha - \frac{4\Omega^2}{3\alpha q^2} + \frac{1}{4 q^6 \sqrt{2 A_1 - 1/(2q^2)}}},\nonumber
\end{eqnarray}
with
\begin{equation}
\alpha = \frac{4 (\frac{2}{3})^{\frac{1}{3} }\Omega^2}{(9 \Omega^2 + \sqrt{81 \Omega^4 - 748 q^6 \Omega^6})^{\frac{1}{3}}}.\nonumber
\end{equation}
It is straightforward to check that the expression (\ref{eq:lrn}) reduces to  (\ref{eq:eos}) in the limit $q\rightarrow 0$. Near the bifurcation point, the behavior of the angular momentum as a function of $\Omega$ for fixed charge and various values of energies is shown in figure (\ref{fi:lom}).
 We are now in a position to comment on  various  scaling behaviors and the critical exponents. While our general analysis in section-(\ref{two}) tells us that the exponents $\delta, \gamma$ will take the same values $3$ and $1$  respectively, for $\beta$ we  need to follow similar steps taken earlier for the
 the Schwarzschild black hole, see (\ref{eq:beta}). Choosing $q = 0.4$, we find  $\beta =1/2$. For other values of the charge, we have not checked explicitly  but we do not see  any reason for the value of $\beta$ to be different.

\section{ISCOs and AdS/CFT} \label{three}

Following the discussion on fixed points and ISCOs in various backgrounds, we now focus our attention to asymptotically AdS spacetimes. We explore how the bulk dynamical features (e.g.,  ISCO and the the critical behavior near  it)  are related to  the behavior of certain operators in the boundary gauge theory~\cite{Fidkowski_2004,PhysRevD.79.064016,Kinoshita:2023hgc,Ceplak:2024bja,Hashimoto:2023buz}.  In empty AdS, these orbit states  correspond to a family of  light double twist operators $[{\mathcal O}_L,{\mathcal O}_L]_{n,J}$ which carry spin $J$ (associated with the angular momentum $L$ ) and potentially, other quantum numbers $n$. If we assume that one of these operators is heavy i.e., certain heavy-light double twist operators $[{\mathcal O}_H,{\mathcal O}_L]_{n,J}$, then the bulk interpretation is in terms of particle orbits in the dual black hole in AdS. The dynamics of such heavy-light operators has been extensively studied earlier using the four point functions in the light-cone bootstrap approach, yielding compelling results. It is believed that the orbit states correspond to narrow resonances, associated with the emergence of quasi-normal modes~\cite{PhysRevD.64.084017}, and the eigenstate thermalization hypothesis~\cite{Srednicki_1999,D_Alessio_2016,lashkari2016eigenstatethermalizationhypothesisconformal,10.21468/SciPostPhys.9.3.034}. \\

\noindent With these insights, in this section, we first compute the binding energy of the stable circular orbits around a five dimensional Schwarzschild blackhole in AdS. It is well known that the 
anomalous dimensions of the double twist CFT operators referred above ~\cite{Berenstein:2020vlp,Sekino:2008he,Abajo-Arrastia:2010ajo,Balasubramanian:2011ur,Nozaki:2013wia,Shenker:2013pqa,Festuccia:2008zx,Fitzpatrick:2014vua,Dodelson:2022eiz} can be extracted from the binding energy of such orbits. This essentially requires one to compute the quantity $(E-L/l -1)$ associated with a circular orbit.  For large $L$ (or equivalently small $m$)\footnote{As in Kepler problem, the relevant parameter here is $L/\sqrt{m}$.}, which translates into anomalous dimension of twist operator with large spin, we compute this quantity in inverse power of spin (or in powers of $m$). Subsequently, stretching the formula for the saddle, we find 
the  leading contribution to the anomalous dimension is of  the form $L/\sqrt{m}$.  Finally, in the limiting case of ISCO, we find 
the anomalous dimension develops non-analytic behavior and becomes complex.  In the concluding subsection we study the behavior of double twist operators in the dual CFT.  After a brief review of the earlier results  about the light-cone bootstrap calculation of the heavy-light four point function, we  find the corrections to the mean filed theory behaviour of these correlators. These corrections, which are purely derived in the boundary CFT, are yet to be interpreted in the bulk. We only offer some suggestions 
at the end of the subsection.

\subsection{Fixed points and ISCOs: Anomalous dimensions} \label{three.1}

In this subsection, we work with a five dimensional AdS-Schwarzschild black hole, setting  $d = 4$ in (\ref{eq:hole}). However, the pattern is  similar for  any $d > 2$. We start with (\ref{eq:xx}) and find the locations of the fixed points. In what follows,  we  differentiate the scaled variables from the rest by keeping the tildes explicitly. Since (\ref{eq:xx}) is a cubic equation in $\tilde x^2$, exact solutions are easily obtained. We find two real, positive solutions - one for the center and the other for the saddle. In the large 
$\tilde L$ limit, these solutions have particularly simple expansions in $1/\tilde L$, 
\begin{eqnarray}
&&\tilde x^{center}_{fp} = \frac{1}{{\tilde l^{\frac{1}{2}} \tilde L^{\frac{1}{2}}}} + \frac{1}{\tilde l^{\frac{3}{2}}\tilde L^{\frac{3}{2}}} + \frac{ 9+ \tilde l^2}{2 \tilde l^{\frac{5}{2}} \tilde L^{\frac{5}{2}}} + \frac{55 + 7 \tilde l^2}{2 \tilde l^{\frac{7}{2}}\tilde L^{\frac{7}{2}}}+  \frac{1547 + 234 \tilde l^2 + 5\tilde l^4}{8\tilde l^{\frac{9}{2}}\tilde L^{\frac{9}{2}}} + {\cal{O}}\Big(\frac{1}{\tilde L^{\frac{11}{2}}}\Big),\label{eq:sadcon}\\
&&\tilde x^{saddle}_{fp} = \frac{1}{2} - \frac{8+\tilde l^2}{2 \tilde l^2 \tilde L^2} - \frac{\tilde l^4 + 80 \tilde l^2 + 576}{4\tilde l^4 \tilde L^4}  - {\cal{O}}\Big(\frac{1}{\tilde L^5}\Big).
\label{eq:censad}
\end{eqnarray}
There are a couple of ways to think about the large $\tilde L$ expansions. We can either imagine holding the black hole mass $m$ fixed and consider (\ref{eq:sadcon}) and (\ref{eq:censad}) as expansions in 
large angular momentum $L$ or, equivalently, we can think of holding $L$ fixed and  making expansions in small $m$.  However, we should note that a small black hole in AdS is 
unstable. This happens when  the black hole has  a mass smaller than $l^2$. We will avoid this by assuming $m/l^2 > 1$. Or, in other words, 
we work with small $l$, smaller than $\sqrt m$ and hence with large black holes in AdS. \\

\noindent Rewriting the large $\tilde L$ expansion as a series in $m$ by converting back to the unscaled variables , the above expansions reorganize as
\begin{eqnarray}
&&x^{center}_{fp}  = \frac{1}{(lL)^{\frac{1}{2}}} + \Big[ \frac{1}{2 (lL^5)^{\frac{1}{2}}} + \frac{1}{(lL)^{\frac{3}{2}}}\Big] m \nonumber\\
&&\qquad\qquad\qquad +\Big[ \frac{7}{2 (l^3L^7)^{\frac{1}{2}}} + \frac{5}{8(L^9 l)^{\frac{1}{2}}}+ \frac{9}{2 (lL)^{\frac{5}{2}}}\Big] m^2 + {\cal{O}}(m^3),\label{eq:sadcon1}\\
&&x^{saddle}_{fp} = \frac{1}{2 m^{\frac{1}{2}}} - \frac{m^{\frac{1}{2}}}{2 L^2} - \Big[ \frac{1}{4 L^4} + \frac{4}{l^2 L^2}\Big] m^{\frac{3}{2}} + 
{\cal{O}}(m^{\frac{5}{2}}).
\label{eq:sadcon2}
\end{eqnarray}
For subsequent analysis, we will require the expression of energy. It follows from  (\ref{eq:leo})
\begin{equation}
E^2 =\frac{ [l^2 x_{fp}^2 (1 - 2 \mul x_{fp}^2) +1]^2}{l^4 x_{fp}^4 (1 - 4 \mul x_{fp}^2)}.\nonumber
\label{eq:elo}
\end{equation}
For small $m$, $E$ has the following series solution:
\begin{equation}
E = 1 + \frac{L}{l} - \frac{m}{lL}  -\frac{(l+ 4 L)m^2}{2l^2L^3} + {\cal{O}}(m^3).
\end{equation}
The above expression allows us to compute the anomalous dimension $\gamma$ of the corresponding
heavy light twist operator on the boundary.  Following ~\cite{Dodelson:2022eiz}, we get
\begin{equation}
\frac{\gamma}{\Delta_L} = E - \frac{L}{l} -1 = - \frac{m}{lL} -  \frac{(l+4L)m^2}{2L^3 l^2} + {\cal{O}}(m^3).
\end{equation}
Here $\Delta_L$ is the scaling dimension of the operator representing the mass of the probe in the bulk.
As for the saddle, we can perform a similar computation, to obtain
\begin{equation}
\frac{\gamma}{\Delta_L} =\frac{L}{2\sqrt{2\mul}}  - \Big(1 +  \frac{L}{l}\Big) + \frac{1}{\sqrt{2}}\Big(
\frac{1}{L} + \frac{2L}{l^2}\Big) \sqrt{\mul} + \frac{(l^4 + 32 l^2 L^2 - 32 L^4) m{\sqrt{m}}}{8\sqrt{2}l^4 L^3}
+ {\cal{O}}(m^2 {\sqrt m}).
\label{eq:anosad}
\end{equation}
A few comments on (\ref{eq:anosad}) are now  in order. We  observe that the constant term $-(1 + L/l)$ represents the energy difference even without a black hole ($\mul=0$), i.e., the energy of the background AdS and is expected. The physical origin of the divergent $\mul^{-1/2}$ term is easy to see if we interpret small mass expansion as referring to the small black holes. First note that for small mass black hole, the Schwarzschild radius is $r_{h} \approx \sqrt{2\mul}$ and the photon sphere (for sufficiently large angular momentum, photon sphere always exists) radius is  $r_{\text{PS}}= 2\sqrt{\mul}$. As $\mul \to 0$, the unstable orbit  radius  $r \sim r_{\text{PS}}  \sim 2\sqrt{\mul} \to 0$ as well. At the same time, the gravitational force in five dimensions scales as $\sim 1/r^3$ near the origin  and hence, to maintain the orbit against diverging force requires diverging energy, which explains the leading term in eqn. (\ref{eq:anosad}). \\

\noindent Before concluding this subsection, we note the following. The temperature of the black hole in the limit of small mass $m$ (and with ${\sqrt m} > l$), is given by
\begin{equation}
T = \frac{f^\prime(r_h)}{4\pi} \sim \frac{r_h}{\pi l^2} \sim \frac{\sqrt {2m}}{\pi l^2}.
\end{equation}
 The Hawking-Page temperature, above which the black hole is stable, is given by $T_{HP} = \frac{3}{2\pi l}$. 
The anomalous dimension (\ref{eq:anosad}), expressed in terms of the temperature,  then takes the form
\begin{equation}
\gamma \sim \frac{L}{2{\sqrt{2 m}}} \sim \frac{L}{2\pi l^2 T}.
\end{equation}
One might be able to extract this  $1/T$  behavior  by studying thermal Green's functions which may have poles in the complex plane, corresponding to quasi-normal frequencies. In other words, one expects the unstable orbit to correspond to quasinormal modes with frequencies
$ \omega_{\text{QNM}} \sim \frac{1}{T} \quad \text{as } T \to T_{HP}$. It thus seems that the anomalous 
dimensions coming from the saddle reveal how bulk near-horizon physics ($r \sim \sqrt{\mul}$) generates $1/T$-divergent contributions to thermal 
correlator poles, which should be observable in finite-temperature CFT computations.  We conclude this subsection reemphasising  that our understanding of a saddle   in the CFT  
is yet unclear  and deserves further study. 

\subsubsection{Non-analyticity of Anomalous dimensions at the ISCOs} \label{three.2}
In the previous subsection, we noted the contrasting behavior of anomalous dimensions in the CFT for the fixed points in the bulk corresponding to  center and saddle. We can now try to go one step further and study how the anomalous dimensions behave in the limiting case of ISCOs, with
$V_{\rm eff}=E^2,\,V_{\text{eff}}'(r_{\tiny{c}},L_{c})=0,\, V''_{\rm eff}(r_{c},L_{c})=0$,
where, $V_{\rm eff}$ is identified from eqn. (\ref{Veff}).  To do this, we 
let $r = r_c + \epsilon$ and $L = L_c + \delta L$, with $\epsilon/r_c, \delta L/L_c \ll 1$, and expand $V_{\text{eff}}'(r_c, L)$ around $(r_c, L_c)$ as
\begin{equation}
V_{\text{eff}}'(r_c + \epsilon, L_c + \delta L) \approx 
\underbrace{V_{\text{eff}}'(r_c, L_c)}_{=0} + 
\frac{\partial V_{\text{eff}}'}{\partial r}\epsilon + 
\frac{\partial V_{\text{eff}}'}{\partial L}\delta L + 
\frac{1}{2}\frac{\partial^2 V_{\text{eff}}'}{\partial r^2}\epsilon^2 + \cdots = 0 \, .\nonumber
\end{equation}
At ISCO, only the last two terms are important, and to leading order we get
the scaling $\epsilon \sim \sqrt{\delta L}$.
Now expanding $V_{\text{eff}}''(r)$ around $r_c$,
\begin{equation}
V_{\text{eff}}''(r) = \underbrace{V_{\text{eff}}''(r_c)}_{=0} + V_{\text{eff}}'''(r_c)\epsilon + \frac{1}{2}V''''(r_c)\epsilon^2 + \cdots \, , \nonumber
\end{equation}
and since $\epsilon \sim \sqrt{\delta L}$, the leading term gives:
$
V_{\text{eff}}''(r) \sim V_{\text{eff}}'''(r_c) \epsilon \sim \sqrt{\delta L}\, .
$
For small radial perturbations around a circular orbit, 
the proper-time frequency is 
$
\omega_t \sim \omega_\tau = \sqrt{V_{\text{eff}}''(r)} \sim (\delta L)^{1/4}\, \sim (L - L_{c})^{1/4}\, . 
$
In AdS/CFT, the anomalous dimension $\gamma$ of heavy-light operators is related to the binding energy, and hence, near the ISCO, one expects (with quantum correction from radial excitations (level $n$)) :
\begin{equation}
\gamma_n(L) - \gamma_{\text{cl}}(L) = \left(n + \frac{1}{2}\right)\omega_t(L) + O(\omega_t^2)  \sim \left(n + \frac{1}{2}\right)(L - L_{c})^{1/4} + O(\omega_t^2) +\cdots \nonumber
\end{equation}
\noindent
As anticipated,  we note that for $L > L_{c}$, there are stable circular orbits where real $\gamma(L)$ exist, followed by the case,
$L = L_{c}$, where marginally stability is reached, with $V_{\text{eff}}''(r_c) = 0$.  Now, for the 
case $L < L_{c}$, we have unstable circular orbits\footnote{Note that for 
case $L < L_{c}$, the ISCOs stop existing, as there are no valid solutions to the conditions: $V_{\rm eff}=E^2,\,V_{\text{eff}}'(r_{\tiny{c}},L_{c})=0,\, V''_{\rm eff}(r_{c},L_{c})=0$, which also satisfy the positivity constraints from energy and angular momentum (see figure-1 of~\cite{Paul:2024khi}).}. Here, setting $\epsilon \sim \sqrt{\delta L} = \sqrt{-|\delta L|} = i\sqrt{|\delta L|}$, the frequency can be written as $\omega \sim \sqrt{i|\delta L|^{1/2}} = e^{i\pi/4}|\delta L|^{1/4} $
Unstable circular orbits have a Lyapunov exponent $\lambda = \sqrt{-V_{\text{eff}}''(r_0)} > 0$ and the perturbations grow as $e^{\lambda t}$, giving the decay width as
\begin{equation}
\Gamma = 2\lambda \sim \sqrt{|\epsilon|} \sim |\delta L|^{1/4}\, . \nonumber
\end{equation}
This is reason why in the CFT, the scaling dimension becomes complex, 
as the anomalous dimension acquires an imaginary part, i.e., $ \gamma_{n,L} = \gamma_R(L) + i\gamma_I(L) \quad \text{for} \quad L < L_{c}$, 
where,
\begin{align}
\gamma_R(L) &= \gamma_{\text{ISCO}} - a|\delta L| + \cdots \\
\gamma_I(L) &= -\frac{b}{2}|\delta L|^{1/4} + \cdots \quad (\text{negative}) \, \nonumber
\end{align}
with arbitrary constants $a,b$.
These complex dimensions are expected to appear in the  finite-temperature CFT correlators as poles, connected to the quasinormal modes of the black hole. The imaginary part $\gamma_I < 0$ gives exponential decay $e^{\gamma_I t}$ in the correlator, characteristic of thermal systems.\\

\noindent
We should mention that the non-analyticity of anomalous dimensions at the ISCO found here are compatible with earlier observations of long-lived orbit states terminating abruptly at ISCO, through the computation of energy associated with radial fluctuations about stable circular orbits~\cite{Berenstein:2020vlp}, which were subsequently extended to long-lived resonances/quasinormal modes in~\cite{Dodelson:2022eiz}.  Related non-analytic structures of finite-temperature CFT correlators associated with bulk null geodesics and the photon sphere were studied earlier in ~\cite{HubenyLiuRangamani2007,DodelsonOoguri2021}, but these works did not study the way the heavy-light anomalous dimensions become non-analytic as the ISCO is approached.  What is found here is a characterization of the ISCO endpoint more precisely as the merger of the center and saddle circular orbit branches. In this sense, the earlier expectation of an abrupt finite-spin endpoint is sharpened into a definite critical behavior in the CFT spectrum. Beyond the endpoint, the corresponding continuation is more naturally interpreted in terms of resonance or quasinormal mode widths rather than complex operator dimensions.

\subsection{Orbits in CFT} \label{three.3}
In the last two subsections, our strategy  was to extend the results on fixed point structure of the bulk geodesics via AdS/CFT to the boundary gauge theory and study behavior of anomalous dimensions near the ISCOs. The study of behavior of CFT correlators can be done independently in the boundary CFT itself, as the notion of orbits in an 
AdS-Schwarzschild background can actually be set up from a CFT point of view. The bound orbits in the bulk can be understood in terms of the spectrum of states in the CFT, at large spin. The large spin expansion has been explored in the last decade in ~\cite{Berenstein:2020vlp,Sekino:2008he,Abajo-Arrastia:2010ajo,Balasubramanian:2011ur,Nozaki:2013wia,Shenker:2013pqa,Festuccia:2008zx,Fitzpatrick:2014vua,Dodelson:2022eiz}, and the connection of orbits to the double-twist operators has been presented in ~\cite{Berenstein:2020vlp,Festuccia:2008zx,Dodelson:2022eiz}. The stress tensor two-point function defines the central charge $c_T$, which can be thought of as the degrees of freedom of a given CFT. We take a pair of operators to be heavy $\Delta_H \approx O(c_T)$, this creates a classical black hole background as we take $c_T \to \infty$ keeping $\mu' = \frac{\Delta_H}{c_T}$ fixed.  The second pair of operators are taken to be light $1 \ll\Delta_L \ll c_T $. We use them to probe the background created by the heavy operators. Using the usual AdS/CFT dictionary, the boundary angular momentum of the operators is defined as $J':= \frac{\Delta_L}{l}\, L$. We have used conformal symmetry to put all four operators in a 2D plane. The orbit states correspond to the double-twist operators in the CFT, 	$\Or_H \Box^n \partial_{\mu_1} \dots \partial_{\mu_{J'}} \Or_L$
		with dimensions $\Delta(n',J')$ given by~\cite{Berenstein:2020vlp,Dodelson:2022eiz}
		\beq
		\Delta(n',J') = \Delta_{\Or_H} +
		\Delta_{\Or_L} + 2n' + J' 	+ \gamma_{n'} (J')
		\eeq 
		that are perturbatively stable in $\frac{1}{J'}$.  The anomalous dimensions in the CFT correspond to binding energy between the black hole and the probe in the bulk~\cite{Fitzpatrick:2014vua, Kulaxizi:2018dxo}.\\
		
\noindent		In the large spin limit $(J' \to \infty)$, the relevant states are perturbatively stable as discussed in ~\cite{Berenstein:2020vlp,Dodelson:2022eiz}. The states manifest themselves as poles in retarded thermal correlation function $G_R (\omega)$, to be discussed below and are also called the quasi normal modes in the bulk dual. They are related to each other via the ETH. It is understood that they have  a  non-perturbative part in spin, due to the tunneling of the orbits into the black hole. The imaginary part, is an exponentially small quantity of the order of $e^{-(J')}$, indicating a long time-scale before the state thermalizes ~\cite{Festuccia:2008zx}. In ~\cite{Dodelson:2022eiz}, the lifetime of stable orbits in the case of large black holes, $\mu' \gg 1, $ and large spin $J \gg \mu' \gg 1$ were found to be of the order $\frac{J'^4}{\mu'^8} \,c_T$ in four dimensions. This is the reason for the state being only meta-stable, but taking the large $c_T $ limit, the effect due to radiation is suppressed. 
		We review the computations and arguments used to relate the stable orbits in AdS, to the double-twist operators in a holographic CFT.
		We define the four point function as follows 
		\beq
		G(z,\bar{z}) =  \expect{\Or_H(0,0) \, \Or_L(z,\bar{z}) \, \Or_L(1,1) \, \Or_H (\infty)} 
		\eeq
		\beq
		u =\frac{x_{12}^2 x_{34}^2}{x_{13}^2 x_{24}^2} = z\bar{z} \hspace{1cm} v = \frac{x_{14}^2 x_{23}^2}{x_{13}^2 x_{24}^2} = (1-z)(1-\bar{z}).
		\eeq		
		We denote $\Or_H(\infty) = \lim_{x_4 \to \infty} |x_4|^{\Delta_H} \Or_H (x_4)$ to scale out the dimensional dependence of $x_4$ in the four point function.\\		
		
		\noindent In the heavy-light channel, the four point function is expanded as, 
		\begin{align}
			\text{s-channel} :\hspace{0.2cm} G(z, \bar{z}) = &(z \bar{z})^{-\half (\Delta_H + \Delta_L) }  \non \\
			& \times\, \sum_{\Or_{\Delta', J'}}|\lambda_{H,L,\Or_{\Delta', J'}}|^2 g_{\Delta,J'}^{\Delta_{H,L}, -\Delta_{H,L}} \, (z, \bar{z}) \hspace{0.5cm} |z| < 1,
		\end{align}
		where $\Delta_{H,L} \equiv \Delta_H-\Delta_L$ and $\lambda_{H,L,\Or_{\Delta, J}}$ are the OPE coefficients. 
		Likewise, in the t-channel we have 
		\begin{align}
			\hspace{1cm}\text{t-channel} :\hspace{0.2cm} &G(z, \bar{z}) = ([1-z][1-\bar{z}])^{-\Delta_L }  \non \\
			& \times\, \sum_{\Or_{\Delta, J}} |\lambda_{L,L,\Or_{\Delta, J}}|  \, |\lambda_{H,H,\Or_{\Delta, J}}|g_{\Delta,J}^{0,0} \, (1-z, 1-\bar{z}) \hspace{0.5cm} |1-z| < 1.
		\end{align}
		Consider the s-channel where the spectrum of exchange operators are the heavy-light double twist ones.  
		\begin{align}
		 G(z, \bar{z}) &= (z \bar{z})^{-\half (\Delta_H + \Delta_L) }  \non \\
			& \times\, \sum_{\Or_{\Delta, J}}|\lambda_{H,L,\Or_{\Delta, J}}|^2 g_{\Delta,J}^{\Delta_{H,L}, -\Delta_{H,L}} \, (z, \bar{z}) \hspace{0.5cm} |z| < 1,
		\end{align}
		where $g_{\Delta,J,}^{\Delta_{H,L}, -\Delta_{H,L}} \, (z, \bar{z})$ are the relevant conformal blocks found in the literature. We consider the limit where $\Delta_H$ and the central charge of the CFT, $c_T$ are large. In this limit, the descendants are suppressed ~\cite{Jafferis:2017zna} and we obtain an expression of the form ~\cite{Dodelson:2022eiz}, 
		\begin{align} \label{ETH4pt}
			G(z, \bar{z})  
			= \sum_{J'=0}^\infty \int_{-\infty}^{\infty} \, d\omega \, |\lambda_{H,L,\Or_{\Delta, J'}}|^2\, (z\bar{z})^{\frac{\omega-\Delta-J'}{2}} \,\,\frac{z^{J'+1}-\bar{z}^{J'+1}}{z - \bar{z}}.
		\end{align}
		We can now relate the heavy-light four point function, and the thermal two-point function~\cite{Dodelson_2023} as follows, 
		\beq
	|\lambda_{H,L,\Or_{\Delta, J'}}|^2 \, = \frac{J'+1}{(\Delta_L -1)(\Delta_L-2)} \frac{\text{Im} \,G_R ({\omega, J'})}{1-e^{-\beta \omega}}.
		\eeq
		Thus, perturbatively, Im $G_R (\omega)$ becomes a sum over the delta functions $\delta(\omega - (\Delta_L+ 2n + J'))$. The term $\frac{1}{1-e^{-\beta\, \omega}}$, for $\omega \approx J \gg 1$ becomes a theta function, provided we ignore the non-perturbative contributions of the order, $e^{-(J)}$. In this way, 	$|\lambda_{H,L,\Or_{\Delta, J}}|^2$ gives us the spectrum of double twist states, $\Or_H \Box^n \partial_{\mu_1} \dots \partial_{\mu_n} \Or_L$ at large spin as expected from light-cone bootstrap\footnote{It would be interesting to address the question of non-perturbative in spin effects due to the presence of the horizon using the Lorentzian Inversion formula~\cite{Caron_Huot_2017} and dispersion relations in the CFT~\cite{Caron_Huot_2021}.}.
		
		\subsubsection{Corrections to MFT behavior} \label{three.4}
		We present a brief computation with regards to corrections to MFT coefficients where in the light-cone bootstrap approach, it is assumed the heavy state contributions are ignored. We will consider the $\frac{1}{\Delta_H}$ contributions to the same and estimate the corrections to the anomalous dimensions. 
		We aim to study the behavior of ISCO's in the dual CFT regime.  We require certain modifications to the light-cone bootstrap problems to study  the behavior of the orbits as we approach the region of the order of event horizon\footnote{We remember that $r_{BH} \approx \sqrt{2 m}$ and $r_{isco} \sim O(3r_{BH}).$}.  
		The mean-field OPE coefficients ~\cite{Komargodski:2012ek} are as follows
		\begin{align}
			|\lambda_{H,L,\Or_{\Delta, J}}|^2 \equiv c^{\text{free}}_{n',J'} &\;=\;
			\frac{
				(\Delta_H + 1 - \tfrac{d}{2})_{n'}\,
				(\Delta_L + 1 - \tfrac{d}{2})_{n'}\,
				(\Delta_H)_{n'+J'}\,
				(\Delta_L)_{n'+J'}
			}{
				n'!\,J'!\,
				(\Delta_H + \Delta_L + n' + 1 - d)_{n'}\,
				(\Delta_H + \Delta_L + 2n' + J' - 1)_{J'}
			}
			\; \non \\
			&  \;\;\;\;\; \times \frac{1}{
				(\Delta_H + \Delta_L + n' + J' - \tfrac{d}{2})_{n'}\,
				(J' + \tfrac{d}{2})_{n'}
			}
		\end{align}
	
		This reproduces the identity operator in the t-channel. We then take the heavy limit $\Delta_H \gg \Delta_L , n' , J'$. 
		To leading order\footnote{We repeatedly use the large-(x) asymptotic of the Pochhammer symbol:
			\beq
			(x+a)_k = \frac{\Gamma(x+a+k)}{\Gamma(x+a)}
			\xrightarrow[x\to\infty]{}
			x^k\left(1+\frac{k(2a+k-1)}{2x}+O(x^{-2})\right).
			\eeq 
		}
		in $(1/\Delta_H)$
		
		\beq
		c^{\text{free}}_{n',J'}
		\xrightarrow[\Delta_H\to\infty]{}
		\frac{
			(\Delta_L + 1 - \tfrac{d}{2})_{n'}
			(\Delta_L)_{n'+J'}
		}{
			n'!J'!
			(J' + \tfrac{d}{2})_{n'}
		}
		+
		\mathcal O\left(\frac{1}{\Delta_H}\right)
		\eeq
		The expansion, considering first order correction to the Mean field theory behaviour in the heavy limit is as follows: 
		\beq \label{sf10}
		c^{\text{free}}_{n',J'}=		 	
		c^{(0)}_{n',J'}
		\left[
		1+\frac{1}{\Delta_H}c^{(1)}_{n',J'}
		+O\left(\frac{1}{\Delta_H^{2}}\right)
		\right]
		\eeq
		In terms of Gamma functions
		\beq {\label{sf9}}
		c^{(0)}_{n',J'} =
		\frac{
			\Gamma\left(\Delta_L + 1 - \tfrac{d}{2} + n'\right)
			\Gamma \left(\Delta_L + n' + J'\right)
			\Gamma \left(J' + \tfrac{d}{2}\right)
		}{
			\Gamma\left(\Delta_L + 1 - \tfrac{d}{2}\right)
			\Gamma(\Delta_L)
			\Gamma\left(J' + \tfrac{d}{2} + n'\right)
			\Gamma(n'+1)\Gamma(J'+1) 
		} + \mathcal O\left(\frac{1}{\Delta_H}\right)
		\eeq
				The first correction term $c^{(1)}_{n',J'}$ to the MFT coefficents is as follows
		\beq
		c^{(1)}_{n',J'} = 
		\frac{1}{2}\Big[
		n'(n'-1)
		-2n'(\Delta_L+J')
		-J'(J'-1)
		\Big] = -\frac{1}{2}\Big[
		(2n'+J')(2\Delta_L+J'-1)		 	
		- n'(n'+1)
		\Big]
		\eeq
		We briefly explain our motivations for considering first order correction to the mean-field theory coefficients. As explored in ~\cite{Dodelson:2022eiz}, considering the OPE coefficients up to $c^{(0)}_{n',J'}$, we obtain the double twist operators that are exchanged in the s-channel which reproduces the identity operator in the limit $J \to \infty$. In this limit ~\cite{Dodelson:2022eiz}, these operators precisely correspond to the orbit states in the bulk dual, and they represent the spectrum of heavy-light double twist operators $\Or_H \Box^n \partial_{\mu_1} \dots \partial_{\mu_J} \Or_L$ CFT states as implied by lightcone bootstrap $(z\ll 1- \bar{z} \ll 1)$. However, we are interested in probing the regime deep inside the AdS spacetime (the ISCO region) and we also assume that $\Delta_L\gg1$, for geodesic approximation to hold. We are also motivated by the gravitational self force corrections~\cite{Barack:2022pde} in understanding the ISCO orbits, and scattering states in AdS~\cite{fitzpatrick2011scatteringstatesadscft, Berenstein:2019tcs, Berenstein:2020vlp}
		as we approach the regime of strong-field interactions. So it is important to take into account the modification of geodesic equation itself, as the ISCO analysis relied on the test particle approximation. When the particle has a finite size , there will be significant modification to the geodesic motion. This modification is  accounted for by taking into account the self-force corrections which are of the order, $\epsilon=\frac{m_{probe}}{\mul}$, where $m_{probe}$ is the mass of the probe, and $\mul$ is the mass of the black hole and $m_{probe} \ll \mul$. In our setup, we would consider $\frac{\Delta_L}{\Delta_H}$ expansion. To access these contributions in the CFT, it is important to consider the sub-leading order in the mean field theory  behavior as well. We will see below, this matches with expectations from anomalous dimensions considerations. \\

\noindent		
We also note that the critical angular momentum which supports marginal orbits is of the order, $J'_{min} \approx \Delta_L \mu'^2$ for $\mu'\gg1$, so that we have large AdS black holes that dominate the canonical ensemble. Thus, we can take the large J limit in the above   
				 equation \eqref{sf10} as well.
		Below we treat large spin $(J'\gg{n',\Delta_L,d})$ systematically, first for the leading heavy–limit coefficient $(c^{(0)})$ and then including the $(1/\Delta_H)$ sub-leading correction. 
				Collecting all the terms to leading order,	 
		\beq		 
		c^{(0)}_{n',J'} =
		\mathcal N_{n'}
		J'^{\Delta_L-1}
		\left[
		1
		+\frac{1}{2J'}
		\Big(
		(\Delta_L+n'-1)(\Delta_L+n')	 	
		- n'(d-1)
		\Big)
		+O(J'^{-2})
		\right]
		\eeq
		with the (J')-independent normalization as
		\beq
		\mathcal N_{n'} = 
		\frac{
			\Gamma\left(\Delta_L + 1 - \tfrac{d}{2} + n'\right)
		}{
			\Gamma\left(\Delta_L + 1 - \tfrac{d}{2}\right)
			\Gamma(\Delta_L)\Gamma(n'+1)
		}
		\eeq
		At leading order, the spin dependence is universal:
		$
		c^{(0)}_{n',J'} \sim J'^{\Delta_L-1},
		$
		as expected for double-twist operators in the lightcone/large-spin regime. 
		We next consider the subleading $(1/\Delta_H)$  correction of \eqref{sf10} at large (J') which gives us information regarding the corrections to MFT coefficents. This would correspond to the existance of corrections to the anomalous dimensions in the s-channel, due to $\frac{1}{\Delta_H}$ terms.
			Considering the large-J' expansion of the correction term we expand in powers of J':	 
		\beq
		C^{(1)}_{n',J'}=
		-\frac{1}{2}J'^2
		-\left(n'-\frac{1}{2}\right)J'		 	
		-n'\Delta_L
		+\frac{1}{2}n'(n'-1)
		+O(J'^{0})
		\eeq	 
		Combining large J' structure along with the $\frac{1}{\Delta_H}$ term gives
		\beq
		c^{\text{free}}_{n',J'}=
		\mathcal N_{n'}
		J'^{\Delta_L-1} \left[
		1
		-\frac{J'^2}{2\Delta_H}
		-\frac{\left(n'-\tfrac12\right)J'}{\Delta_H}  +\frac{a_1}{J'}
		+O\left(\frac{J'^0}{\Delta_H}\right)	\right] 		
		\label{heavyMFT}
		\eeq		 
		where
		$$
		a_1=\frac12\Big[(\Delta_L+n'-1)(\Delta_L+n')-n'(d-1)\Big].
		$$
		We notice here, in addition to the $J'^{\Delta_L-1}$ which corroborates with the lightcone bootstrap results, we also obtain an additional term of the form $-\frac{J'^2}{\Delta_H}$ and $-\frac{(n'-\frac{1}{2})J'}{\Delta_H}$ which includes the corrections to the lightcone bootstrap results. By normalizing \eqref{sf9},
		\beq
		c^{\text{free}}_{n',J'} =   \frac{
			\Gamma\left(\Delta_L + 1 - \tfrac{d}{2} + n'\right)
			\Gamma\left(\Delta_L + n' + J'\right)
			\Gamma\left(J' + \tfrac{d}{2}\right)
		}{
			\Gamma\left(\Delta_L + 1 - \tfrac{d}{2}\right)
			\Gamma(\Delta_L)
			\Gamma\left(J' + \tfrac{d}{2} + n'\right)
			\Gamma(n'+1)\Gamma(J'+1) \, 
		} \, \tilde{c}^{\text{free}}_{n',J'}
		\eeq
		such that $\tilde{c}^{\text{free}}_{n',J'} = 1$, we obtain the identity operator in the t-channel using the crossing equation. As a check, we reproduce the sub-leading results of anomalous dimensions to lightcone bootstrap.  A single stress tensor exchange in the t-channel  would give us the anomalous dimensions as we expect for the double- twist operators (or the orbit states) ~\cite{Li:2020dqm}
		\beq \label{last}
		\tilde{c}^{\text{free}}_{n',J'} , \gamma_{n', J'} \approx  \frac{1}{J'^{\frac{d-2}{2}}} \non
		\eeq
		This is the usual spectrum of states we obtain from the lightcone bootstrap at large spin. Essentially, this is the $J'^{\Delta_L-1}$ behavior we obtained in the above analysis. 
		We now consider, the sub-leading $\frac{1}{\Delta_H}$ corrections to \eqref{sf10}, by considering the sub-leading heavy terms in the MFT coefficients as obtained in \eqref{heavyMFT}.  While the identity term in the expansion would give us the usual $\gamma_{n', J'}$, including the $\frac{1}{\Delta_H}$ corrections, give us a term that goes as $\frac{J'^2}{\Delta_H} \text{ and } \frac{J'}{\Delta_H}$.\footnote{Although the $J'$ and $J'^2$ terms appear to be at the same order in $\frac{1}{\Delta_H}$, in order to reproduce the identity operator in the dual t-channel, the $J'$ is leading  compared to the $J'^2$ term.} By substituting in the anomalous dimension calculations to leading order, we obtain a positive  sign term in  the anomalous dimension computation.  Note the negative sign in \eqref{heavyMFT} which corresponds to mean field theory coefficients. For the case $n'=0$ corresponding to circular orbits in the bulk, we obtain a leading order solution of the form: 
		\beq \label{last1}
		{\gamma_{n'=0, \,J'}} \approx \frac{J'^2}{\Delta_H} +\frac{J'}{\Delta_H} + \, \Or\left({\frac{1}{\Delta_H}}\right)^2 +0 - \frac{\Delta_H}{c_T} \frac{1}{J^{\frac{d-2}{2}}} +  \Or\left({\frac{\Delta_H}{c_T}}\right)^2
		\eeq 
		Here, 0 refers to mean field theory behavior, this means that the binding energy is zero (or in absence of the black hole), and the expansion is in terms of $\mu' \sim \frac{\Delta_H}{C_T}$, as is usual in the bootstrap literature. We expect here that the $\frac{1}{\Delta_H}$ correction terms in the anomalous dimensions are on account of the radiation reaction corrections of the orbits calculated from the deviation to the geodesic equation due to emission of gravitational waves. This motivates us to study the self-force corrections to bulk orbits. We expect to report on the results of this investigation in the near future.\\

\noindent
We conclude this section by delineating our results with earlier work.  Earlier works established the interpretation of Regge trajectories of heavy-light double-twist operators as orbits in the bulk and related their anomalous dimensions to orbital binding energies \cite{Berenstein:2020vlp,Dodelson:2022eiz}.  They also emphasize that the large-spin description should eventually be extended toward finite spin, where long-lived orbit states can evolve into resonant or quasinormal-mode behaviour.  On the CFT side, finite-spin anomalous dimensions have indeed been obtained earlier using Lorentzian inversion methods \cite{Li:2020dqm,CardonaSen2018,CardonaGuhaKanumilliSen2019}, while higher-order HHLL/Regge analyses have developed corrections in the gravitational-strength parameter~\cite{Kulaxizi:2018dxo,Karlsson:2019qfi}:
$
\mu' \sim \frac{\Delta_H}{C_T} \, .
$
These results provided us a clear motivation for going beyond the asymptotic large-spin/geodesic regime, especially to study the structure close to ISCO's in the bulk.\\

\noindent
The present analysis, in this section  \ref{three.4}, isolates a subleading contribution through the $1/\Delta_H$ expansion of the mean-field OPE data.  Three points are worth emphasizing. First, (\ref{last}) exhibits explicit $1/\Delta_H$ corrections to the heavy-limit MFT coefficients, distinct from the usual expansion in $\Delta_H/C_T$.  Second, these corrections generate the positive $J'^2/\Delta_H$ and $J'/\Delta_H$ contributions appearing in (\ref{last1}), which can compete with the standard negative large-spin binding term as one moves away from the asymptotic lightcone regime and toward the strong-field/ISCO region. Third, this structure suggests a systematic CFT route to corrections beyond the test-particle geodesic approximation.  In particular, as we mentioned above, it motivates us to investigate whether the large but finite-$\Delta_H$ corrections encode conservative self-force effects, while dissipative radiation-reaction physics should instead be associated with finite widths or decay of the corresponding orbit states.  Although the need to go beyond large spin and the appearance of finite-spin resonant physics were anticipated in the literature cited above, the explicit $1/\Delta_H$ structure and its possible connection to the approach to the ISCOs, do not appear to have been isolated previously in the form obtained here.  We are investigating the effect of these corrections to the bulk and hope to report in the near future.

\section{Conclusions} \label{four}

We studied the trajectories of massive particles in general spherically symmetric black holes in arbitrary dimensions, and found certain universal features based on the topological classification of the fixed points. In particular, we found that if the system admits a center, there are two possible outcomes: regardless of
the value of the angular momentum, the center always survives, a possibility realized in global AdS space-time or, the center disappears below a critical value of angular momentum, which happens for various black holes. For the latter case, we found that regardless of the details of the black hole, there must always be a saddle point. Topological arguments show that there exists a certain critical value of energy, angular momentum and the angular velocity, where the center and the saddle coalesce, which happens at a special point in the moduli space, where the trajectories are the  limiting innermost stable circular orbits (ISCOs). At the critical point, the conserved quantities showed scaling behavior reminiscent of a second order phase transition, whose mean field exponents are universal and match the ones from the van der Waals system. We presented several examples of massive particles in Schwarzschild and Reissner-Nordstrom black holes, and explicitly computed the critical exponents, which are valid for general backgrounds. \\

\noindent 
Following the discussion of fixed points and ISCOs in general backgrounds in the bulk, we focused our attention on asymptotically AdS backgrounds. It is known that from AdS/CFT correspondence, the anomalous dimension $\gamma$ of certain CFT operators can be extracted from the energy and angular momentum of the orbits. Following this logic, in section-(\ref{three}), using a certain scaling regime,  for the center, we recovered the usual results giving a negative value of $\gamma$ for certain double twist operators in the large spin limit, where as, for the saddle, we find that $\gamma$ comes out to be positive. On the CFT side, we studied the behavior of the center, and as for the saddle solution, we made some speculations on the possible interpretation of its anomalous dimensions, which needs further exploration. From the bulk saddle solution, we obtained a correction term to $\gamma$ as  $\frac{L}{2\sqrt{m}}$, whose interpretation in the boundary is unclear, and needs further investigation. While far away from the black hole, the lightcone bootstrap results and Regge limit at large impact parameter overlap, however, as we approach the ISCO, the behavior of the anomalous dimensions is controlled by the Regge limit. The thermal two-point correlator in the boundary CFT, which encodes information about the bulk quasi-normal modes should be a viable diagnostic to understand the saddle fixed point. However, the full connection is not clear to us at present. As mentioned earlier, we expect the $\frac{1}{\Delta_H}$  corrections to MFT behavior to be related to the radiation-reaction and self-force effects in the bulk. The $\frac{1}{\Delta_H}$ corrections have been previously studied in the literature in reference to the Tauberian theorems~\cite{Qiao_2017tt, Mukhametzhanov_2019}, where we are rather motivated to understand the OPE coefficients as being averaged over a small window of states. (In our case, these would correspond to the black hole microstates.) We also saw that as one approaches the ISCO, the anomalous dimensions develop non-analytic behavior and become complex. In~\cite{Jia:2026pmv}, such studies were performed to investigate the scattering behavior at small impact parameters. It was found that the phase shift develops complex values. It would be interesting to understand the connection of~\cite{Jia:2026pmv} with the present work.  \\

\noindent
Near the ISCO, the relativistic nature of the particle would imply that the anomalous dimensions of the operators are no longer of the $O(\frac{1}{J})$. Due to the non-perturbative AdS/CFT correspondence, these quantities are related to the Regge limit at finite impact parameter of the heavy-heavy-light-light four point function of scalar operators in the dual CFT. In~\cite{Kulaxizi:2017ixa}, this Regge limit was studied and it was shown how to extract the phase shift from the eikonal limit of AdS scattering. It has been understood that the phase shift, computed in Gravity is related to the anomalous dimensions~\cite{Kulaxizi:2018dxo},
encoding the information about the deflection angle and time delay. It is interesting to speculate to what extent the phase shift analysis would allow us to extract the anomalous dimensions of the ISCO orbits. As an another avenue for future research, it would be good to understand the scattering formalism in AdS~\cite{fitzpatrick2011scatteringstatesadscft, Paulos_2017, Komatsu_2020} and connect it to bound orbit states that we have studied here, as has been done in flat space. One would have to consider scattering states in AdS-Schwarzschild~\cite{Berenstein:2020vlp,Berenstein:2019tcs}, and consider strong-field effects near the ISCO, which would allow us to extract the corresponding deflection angle and time delay associated to a timelike geodesics in presence of a black hole. Such studies have been carried out in flat-space, and it would be interesting then, to understand the flat space limit of AdS~\cite{Hijano_2019, Li:2021snj}, in which these computations are expected to be equivalent.  

\section*{Acknowledgements}
The work of C.B. is supported by ARG-MATRICS grant no. ANRF/ARGM/2025/002280/MTR. P.C. acknowledges the lecturers and participants of the Kavli Asian Winter School 2025 for stimulating discussions. S.M. thanks  Chethan N. Gowdigere for useful discussions during early stages of this work. We also thank the anonymous referee for helpful suggestions. 

\renewcommand{\thesection}{\Alph{section}}


\bibliographystyle{apsrev4-1}
\bibliography{isco140926.bib}
\end{document}